\documentclass[preprint]{elsarticle}

\usepackage{bm}
\usepackage{graphicx}
\usepackage{amsmath}
\usepackage{leftidx}
\usepackage{stmaryrd}  %Double square parentesys

\newcommand{\capstar}{\overset{*}\cap}%
\newcommand{\deltastar}{\overset{*}\delta}%
\newcommand{\Deltastar}{\overset{*}\Delta}%
\newcommand{\pstar}{\overset{*}{\mathcal P}}%
\newcommand{\pbar}{{\mathcal P}}%
\newcommand{\sump}[2]{{\sum_{#1}}\raise4pt\hbox{\hskip#2$'$}}

\begin{document}
\title{Replica Fourier Transform: Properties and Applications}

\author[sap,smc]{A. Crisanti\corref{c1}}
\ead{andrea.crisanti@roma1.infn.it}

\author[cea]{C. De Dominicis}
\ead{cirano.dedominicis@cea.fr}

\cortext[c1]{Corresponding Author}

\address[sap]{Dipartimento di Fisica, Universit\`a di Roma ``La Sapienza''
             P.le Aldo Moro 2, 00185 Roma, Italy}

\address[smc]{CNR-ISC, 
             Via dei Taurini19, 00185 Roma, Italy}

\address[cea]{CEA Sacaly,
                    Ormes des Merisiers, 91191 Gif-sur-Yvette, France}

\begin{abstract}
The Replica Fourier Transform is the generalization of the discrete Fourier 
Transform to quantities defined on an ultrametric tree. It finds use in conjunction of the 
{\sl replica method} used to study thermodynamics properties of disordered systems 
such as spin glasses.
Its definition is presented in a systematic and simple form and its use illustrated with 
some representative examples.   
In particular we give a detailed discussion of the diagonalization in the Replica Fourier Space 
of the Hessian matrix of the Gaussian fluctuations about the mean field saddle point of spin glass 
theory.  The general results are finally discussed for a generic spherical spin glass model, where
the Hessian can be computed analytically.
\end{abstract} 


\maketitle

\section{Introduction}
The development of the replica method for the study of disordered systems
can be traced back to the middle of 70's  with the introduction
of a mean-field model for spin glasses known since then as 
the Sherrington-Kirkpatrick model \cite{SheKir75,SheKir78}.
The search for its solution in the low temperature spin glass phase
has lead to the introduction and development of 
tools such as the {\sl Replica Method} and the concept of 
{\sl Replica Symmetry Breaking}, that have found applications in a variety of
other fields of the complex-system world, just to cite a few,  
neural networks, combinatorial optimization and glassy physics.

The hallmark of a spin glass state is the presence of a large number of degenerate,
locally stable states.  As consequence of this  two identical ideal replicas
of the system, indistinguishable in the high temperature paramagnetic phase, 
in the spin glass state may lie into different 
states and, hence,  distinguishable.
The symmetry among replicas is  {\sl broken}. 
A solution which accounts for this scenario 
was proposed by Parisi \cite{Parisi79,Parisi80}, and it is 
now accepted as the correct mean-field solution of the Sherrington-Kirkpatrick model,
and its generalizations.
For finite dimensional systems this solution is known to exists at least down to 
dimension $6-\epsilon$ \cite{MooBra11,ParTem12}.
However its physical relevance below the critical dimension $d_c=6$, 
and hence for dimension $3$ or $2$, is still an open problem  and a different scenario
the {\sl droplet} picture \cite{BraMoo86}, an essentially replica symmetric description, 
has been proposed.

According to the Parisi solution in the spin glass phase the replicas in the Replica Space are 
organized on an {\sl ultrametric tree}. In this context the RFT is the generalization of the discrete 
Fourier Transform  to quantities defined on an ultrametric tree and,
similar to the Fourier Transform, the  RFT is powerful tool of the replica method.
It can indeed be used not only for the study of the mean-field solution but also to investigate
finite dimensional corrections \cite{DeDomCar96}.

The use of RFT was introduced with the work of Carlucci, De Dominicis, Kondor and 
Temesv\'ari at the end of 90's
\cite{DeDomCar96,DeDomCarTem97,CarDeDom97,DeDomKonTem98}.
Since then it has been used in several other papers. 
However to our knowledge a systematic and simple presentation of RFT  is still missing. 
This makes difficult to 
enter into the subject even for researcher working in the field of disordered systems.
The aim of this work is fill this missing presenting the use of the RFT,  and its properties,
discussing some representative examples.
The presentation is kept simple so it could serve as a ``reference" to enter into the world of RFT. 

In Sections \ref{sec:RepSpa} and \ref{sec:RFT}  we introduce the notation and give  
the definition of the RFT extending  the discrete 
Fourier Transform to functions defined on an ultrametric tree. 
In Section \ref{sec:ExRFT} the RFT is evaluated for some typical functions occurring in
the study of spin glass models. In Section \ref{sec:EigenEq}  we illustrate the use of RFT to solve 
eigenvalue problems in the Replica Space by discussing the diagonalization of matrices whose 
structure is relevant for the analysis 
of the Gaussian fluctuations about the saddle point in spin glass models. 
Finally in Section \ref{sec:SphMod} we compute the Hessian of the Gaussian fluctuations for a 
generic mean field spin glass spherical model generalizing to an arbitrary number of replica 
symmetry breaking  the result of Ref. \cite{CriSom92}.
Details of some calculations can be found in the Appendixes.

\section{The Replica Space}
\label{sec:RepSpa}
In the replica space each replica is identified by an index $a=0,\dotsc, n-1$, 
where $n$ is the number of replicas.  In view of the introduction of the 
Replica Fourier Transform it is useful to use zero-offset values, and periodic 
boundary conditions, so that each index $a$ is given by the sequence of the 
first  $n$ integers  {\sl modulo} $n$.

The metrics is introduced in the Replica Space by the notion of {\sl overlap}, 
or (co-)distance, between two replicas.
In a scenario with $R$ {\sl replica symmetry breaking steps} the overlap takes 
$R+1$ different  values.
Then by considering an ordered sequence of $R+2$ integer $p_{k} > p_{k+1}$, 
with $p_0 =n$ and $p_{R+1} = 1$, and the requirement that $p_k/p_{k+1}$ is 
integer, and writing $a = \sum_{k=0}^R a_k\, p_{k+1}$, each index $a$ can  be 
mapped to a set of $R+1$ integers $a_k$:
\begin{equation}
  a = [a_0,\dotsc, a_R] 
  \qquad \mbox{with}\qquad 
  a_k = 0, \dotsc, p_{k}/p_{k+1}-1.
\end{equation}
Under the assumption of periodic boundary condition, each index $a_k$ is given 
by the sequence of the first $p_k/p_{k+1}$ integers {\sl modulo} $p_k/p_{k+1}$.

\begin{figure}[h]
\centering
\includegraphics[scale=0.8]{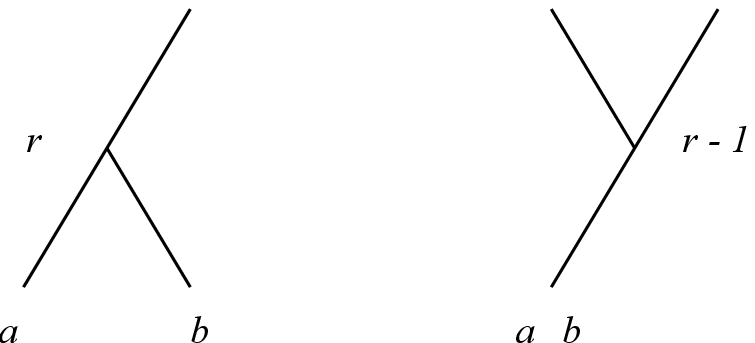}
\caption{Overlap $a\cap b = r$ (left) and co-overlap $a\capstar b = r$ (right).}
\label{fig:overlap}
\end{figure}

Two replicas $a$ and $b$ are said to have and {\sl overlap} $r$, denoted by
$a\cap b = r$, if $a_k=b_k$ for $k=0,\dotsc, r-1$ and $a_r\not=b_r$:
\begin{equation}
  a\cap b  = r  \quad\Rightarrow\quad
  \left(\begin{matrix} a\\ b \end{matrix}\right) = 
  \left[
     \begin{matrix}
  \phantom{x} & \phantom{x} & \phantom{x} & a_{r}' & a_{r+1} & \dotsb & a_{R}\\
 a_0 & \dotsb & a_{r-1} & \phantom{x} & \phantom{x} & \phantom{x} & \phantom{x} \\
  \phantom{x} & \phantom{x} & \phantom{x} & b_{r}' & b_{r+1} & \dotsb & b_{R}\\
      \end{matrix}
  \right]
\end{equation}
where the prime means that $a_r\not= b_r$. Alternatively we can write 
$a-b = [0,\dotsc,0,a_r'-b_r',a_{r+1}-b_{r+1},\dotsc,a_R-b_R]$ or, if we 
introduce the replica $0 = [0,\dotsc,0]$, as $(a-b)\cap 0 = r$. 
The graphical tree representation of the overlap $a\cap b= r$ is shown in 
Fig. \ref{fig:overlap}.

In the Parisi replica symmetry breaking scheme, where the $n\times n$ overlap 
matrix $Q^{ab}$ between replica $a$ and $b$ is divided along the diagonal into 
boxes of decreasing sizes, the integers $p_k$'s are the linear size of the 
boxes. Then $a\cap b= r$ means that replica $a$ and $b$ belong to the same 
Parisi box of size $p_r$ but to two distinct boxes of size  $p_{r+1}$. 
The overlap matrix  $Q^{ab}$ assumes the same value $Q_r$
for all replicas $a$ and $b$ with $a\cap b = r$. 

The overlap divides the Replica Space into a complete set of disjoint shells 
$S_r = \{a,b: a\cap b = r\}$,
identified by the  characteristic function:
\begin{equation}
 \Delta_r(a,b) = \prod_{k=0}^{r-1} \delta_{a_k,b_k}\,(1 - \delta_{a_r,b_r})
               = \delta_r(a,b) - \delta_{r+1}(a,b),
\end{equation}
where $\delta_{x,y}$ is the Kronecker delta and 
\begin{equation}
 \delta_r(a,b) = \sum_{k=r}^{R+1}\Delta_{k}(a,b)
               = \left\{\begin{array}{ll}
                                                       1 & r = 0, \\
       {\displaystyle \prod_{k=0}^{r-1}\delta_{a_k,b_k}} & r = 1,\dotsc, R+1.
                        \end{array}
                 \right.
\end{equation}
The function $\delta_r(a,b)$ for $r=R+1$  reduces to: 
$\delta_{R+1}(a,b) = \delta_{a,b}$.\footnote{
Identities between indexes must be intended modulo their periodicity. Therefore
$\delta_{a_r,b_r}$ means $a_r = b_r \pmod{p_r/p_{r+1}}$ while  
$\delta_{a,b}$ means $a = b \pmod n$ or, equivalently, 
$a_r = b_r \pmod{p_r/p_{r+1}}$ for $r=0,\dotsc, R$. 
}
We adopt the convention that quantities with one or more indexes outside 
the range $[0,R+1]$, for example $\delta_{R+2}(a,b)$, are null. 
This simplify the handling of boundary terms.
The function $\delta_r(a,b)$ defines the subspace 
$B_r = \{a,b : a\cap b \geq r \}$ given 
by the sum of all shells $S_t$ with $t\geq r$:  $S_r = B_r - B_{r+1}$.

\begin{figure}[h]
\centering
\includegraphics[scale=0.8]{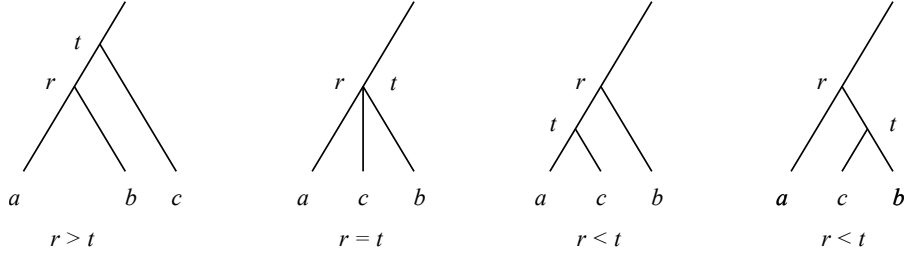}
\caption{Possible orderings of replicas $a$, $b$ and $c$ with 
$a\cap b = r$, $b\cap c = t$.
                From left to right:
                 $a\cap c = c\cap b < a\cap b$;
                 $a\cap c = c\cap b = a\cap b$;
                 $a\cap c > c\cap b = a\cap b$;
                 $a\cap c = c\cap b < a\cap b$.
                 It is always possible to reorder the three replicas so that
                 $x\cap y \geq x\cap z = y\cap z$.
                 }
\label{fig:3replica}
\end{figure}

The metrics based on the overlap defines an ultrametric geometry in the 
Replica Space: for any three  replicas $a,b,c$ we have
$a\cap c \geq \min (a\cap b, c\cap b)$ as can be easily seen in 
Fig. \ref{fig:3replica}.

In the Replica Fourier Space the natural quantity to consider is the 
complementary of the overlap,  the {\sl co-overlap}.  
Two replicas $a$ and $b$ have a  {\sl co-overlap} $r$, denoted by 
$a\capstar b = r$, if $a_k=b_k$ for  $k=r,\dotsc, R$ and $a_{r-1}\not=b_{r-1}$:
\begin{equation}
  a\capstar b = r  \quad\Rightarrow\quad
  \left(\begin{matrix} a\\ b \end{matrix}\right) = 
  \left[
     \begin{matrix}
 a_{0} & \dotsb & a_{r-2} & a_{r-1}'   &\phantom{x} &\phantom{x} &\phantom{x}\\
 \phantom{x} &\phantom{x} &\phantom{x} &\phantom{x} & a_{r} & \dotsb & a_{R}\\
 b_{0} & \dotsb & b_{r-2} & b_{r-1}'   &\phantom{x} &\phantom{x} &\phantom{x}\\
      \end{matrix}
  \right]
\end{equation}
or, equivalently,  $(a-b)\capstar 0 = r$.
The graphical tree representation of the co-overlap $a\capstar b= r$ is shown 
in Fig. \ref{fig:overlap}.
It is easy to verify that  $a\capstar c \leq \max (a\capstar b, c\capstar b)$
or any triplet of replicas.
As the overlap, the co-overlap divides the space into the complete set of 
disjoint shells $\overset{*}S_r = \{a,b : a\capstar b = r\}$,
whose characteristic function is:
\begin{equation}
 \Deltastar_r(a,b) = (1 - \delta_{a_{r-1},b_{r-1}})\, 
                      \prod_{k=r}^{R} \delta_{a_k,b_k}
                   = \deltastar_r(a,b) - \deltastar_{r-1}(a,b),
\end{equation}
where
\begin{equation}
 \deltastar_r(a,b) =\sum_{k=0}^{r} \Deltastar_k(a,b)
                               = \left\{\begin{array}{ll}
                                       1 & r = R+1 \\
    {\displaystyle \prod_{k=r}^{R}\delta_{a_k,b_k}} & r = 0,\dotsc, R.
                                     \end{array}
                           \right.
\end{equation}
and  $\deltastar_0(a,b) = \delta_{a,b} = \delta_{R+1}(a,b)$.
The function $\deltastar_r(a,b)$ selects the subspace 
$\overset{*}B_r = \{a,b : a\capstar b \leq r \}$ equal 
to the sum of  shells $\overset{*}S_t$ with $t\leq r$:
$\overset{*}S_r = \overset{*}B_r - \overset{*}B_{r+1}$.

It is easy to verify the following identities
\begin{equation}
  \Delta_r(a,b)\, \Delta_s(a,b) = \delta_{r,s}\,\Delta_r(a,b)
\end{equation}
\begin{equation}
\label{eq:Deltastar_ort}
  \Deltastar_r(a,b)\, \Deltastar_s(a,b) = \delta_{r,s}\,\Deltastar_r(a,b)
\end{equation}
and
\begin{equation}
\label{eq:sumDelta}
 \sum_b \Delta_r(a,b) = p_r - p_{r+1}.
\end{equation}

\begin{equation}
\label{eq:Deltastar_sum}
 \sum_b \Deltastar_r(a,b) = p_0\left(\frac{1}{p_r} -\frac{1}{p_{r-1}}\right).
\end{equation}
%

%
\if 0
while,
\begin{equation}
 \sum_{c}\,   \Delta_r(a,c)\, \Delta_s(c,b) = \left\{
   \begin{split}
     & (p_u - p_{u+1})\, \Delta_{v}   & r\not=s\\
     & (p_s - 2p_{s+1})\,\delta_s(a,b) + p_{s+1}\,\delta_{s+1}(a,b) & r = s
   \end{split}
   \right.
\end{equation}
where $u = \max(r,s)$ and $v = \min(r,s)$.
\fi

\section{Replica Fourier Transform}
\label{sec:RFT}
The Replica Fourier Transform (RFT) a  generic function $f^a \equiv f(a)$ 
defined on the Replica Space with $R$ replica symmetry breaking steps at
the {\sl momentum} or {\sl wave number} $\hat{\mu}$ is defined as
\begin{equation}
 \mbox{RFT}\left[f(a)\right](\hat{\mu}) = f(\hat{\mu}) 
                     = \sum_{a} f(a)\,\chi(a \hat{\mu})
\end{equation}
where 
 the {\sl kernel} $\chi(a\hat{\mu})$ is 
\begin{equation}
  \chi(a\hat{\mu}) = \prod_{k=0}^{R} \frac{1}{\sqrt{p_k/p_{p+1}}} 
             \exp\left( 2\pi i \frac{a_k \hat{\mu}_k}{p_k/p_{k+1}} \right)
                 = \frac{1}{\sqrt{p_0}} \prod_{k=0}^{R} 
               e^{2\pi i  \frac{p_{k+1}}{p_k} a_k \hat{\mu}_k},
\end{equation}
and the sum is over all replica indexes
\begin{equation}
 \sum_a\ \overset{def}= \prod_{k=0}^R\sum_{a_k=0}^{p_k/p_{k+1}-1}.
\end{equation}
The moment index $\hat{\mu}_k$  of the Replica Fourier Space 
takes $p_k/p_{k+1}$ 
different integer values, which can be taken equal to 
$0,\dotsc, p_k/p_{k+1}-1$, as the index $a_k$ of the Replica Space, 
or, using the periodicity of the exponential, shifted by half-period to
have positive and negative values,  similar to ordinary discrete Fourier 
Transform.
As for ordinary Fourier Transform, different conventions are used in the 
literature to denote the transformed function.
We shall denote the RFT by adding the ``hat" to the index with respect to which 
the RFT is evaluated.  This convention is quite useful when dealing with RFT of 
function depending on more than one replica index.

It is straightforward to verify the following relations
\begin{equation}
 \label{eq:RFTdir}
  \sum_{a_k=0}^{p_k/p_{k+1}-1} e^{2\pi i  \frac{p_{k+1}}{p_k} a_k b_k} =
      \frac{p_k}{p_{k+1}} \delta_{b_k,0},
\end{equation} 
and
\begin{equation}
 \label{eq:sum}
 \sum_a \chi(ab) = \sqrt{p_0}\, \delta_{R+1}(b,0) 
                              = \sqrt{p_0}\, \delta_{b,0},
\end{equation}
which, together with
\begin{equation}
 \label{eq:prod}
  \chi(a+b) = \sqrt{p_0}\,\chi(a)\,\chi(b),
\end{equation}
gives
\begin{equation}
\label{eq:inv}
 \sum_{c} \chi(ac)\chi(-cb) = \delta_{a,b}.
\end{equation}
allowing for the inversion of the RFT:
\begin{equation}
 \label{eq:RFTinv}
     f(a) = \sum_{\hat{\mu}} f(\hat{\mu})\,\chi(-\hat{\mu}a).
\end{equation}

When computing the RFT of functions defined on the Replica Space 
one eventually has to evaluate the RFT restricted to either subspaces
$B_r$ and $\overset{*}B_r$  and/or to shells $S_r$ and $\overset{*}S_r$.
For these cases the following identities hold.
When the RFT is restricted to subspaces $B_r$ or $\overset{*}B_r$,
then
\begin{equation}
\label{eq:deltachi}
  \sum_{c} \delta_r(a,c)\,\chi(cb) = p_r\,\deltastar_r(b,0)\,\chi(ab),
\end{equation}
\begin{equation}
  \sum_{c} \deltastar_r(a,c)\,\chi(cb) = 
              \frac{p_0}{p_r}\,\delta_r(b,0)\,\chi(ab).
\end{equation}
If, instead,  the RFT is restricted to shell $S_r$: 
\begin{equation}
\label{eq:pstar_1}
  \sum_c \Delta_r(a,c)\,\chi(cb) = \pstar_r(b,0)\,\chi(ab)
\end{equation}
where
\begin{equation}
\label{eq:pstar_2}
  \pstar_r(a,b) = p_r\deltastar_r(a,b) - p_{r+1}\deltastar_{r+1}(a,b).
\end{equation}
For $r=R+1$ the index $r+1$ is outside the permitted range
and the second term is missing:
$\pstar_{R+1}(a,b) = p_{R+1}\deltastar_{R+1}(a,b) = 1$.
The function $\pstar_r(a,b)$ satisfies the sum rule
\begin{equation}
\label{eq:pstar_pstar}
  \sum_{c} \pstar_r(a,c)\,\pstar_s(c,b) = p_0\, \delta_{r,s}\, \pstar_r(a,b).
\end{equation}
Finally when the  RFT is restricted to shell $\overset{*}S_r$:
\begin{equation}
  \sum_{c} \Deltastar_r(a,c)\,\chi(cb) = p_0\, \pbar_r(b,0)\,\chi(ab)
\end{equation}
where
\begin{equation}
\label{eq:Pr}
  \pbar_r(a,b) = \frac{1}{p_r}\delta_r(a,b) - 
                       \frac{1}{p_{r-1}}\delta_{r-1}(a,b).
\end{equation}
When $r=0$ the index $r-1$ lies outside the permitted range and 
the second term is missing:
$\pbar_0(a,b) = (1/p_0)\, \delta_0(a,b)= 1/p_0$. 
The function $\pbar_r(a,b)$ obeys the  sum rule
\begin{equation}
  \sum_{c} \pbar_r(a,c)\,\pbar_s(c,b) = \delta_{r,s}\, \pbar_r(a,b).
\end{equation}

\section{Examples of Replica Fourier Transforms}
\label{sec:ExRFT}

In this Section we illustrate the use of RFT  by discussing the RFT of some 
typical functions which occur in the study of spin glass models.

\subsection{One-replica functions}
The simplest case is that of a function $f^a=f(a)$ of  only  one replica index.
The  function $f(a)$ assumes the value $f_r$ whenever $a$ has  
overlap $r$ with a chosen reference replica in the Replica Space.
By choosing the replica ``$0$'' as reference, then  
$f(a) = f_r$ for $a\cap 0 = r$,  with $r=0\dotsc, R+1$, and $f(a)$ 
can be written as:
\begin{equation}
\label{eq:1repfun}
  f(a) = \sum_{r=0}^{R+1} f_r \Delta_r(a,0)
       = \sum_{r=0}^{R+1}\, [f_r - f_{r-1}]\, \delta_r(a,0)
\end{equation}
since shells $S_r$ form a complete decomposition of the Replica Space.
The second equality follows because quantities with index outside the range 
$[0,R+1]$ are null.
Then, by using Eq. (\ref{eq:deltachi}), the RFT is readily computed and reads:
\begin{equation}
\label{eq:RFTvec}
\begin{split}
  f(\hat{a}) =  \sum_a f(a)\,\chi(a\hat{a})
            &= \chi(0)\, \sum_{r=0}^{R+1} p_r\,[f_r - f_{r-1}]\,
                 \deltastar_r(\hat{a},0)\\
            &= \frac{1}{\sqrt{p_0}}\sum_{r=0}^{R+1} f_{\hat{r}}\,
                 \Deltastar_r(\hat{a},0)
                 \end{split}
\end{equation}
where we have used $\chi(0) = 1/\sqrt{p_0}$, and defined $f_{\hat{r}}$ as:
\begin{equation}
\label{eq:1rep}
   p_r\,[f_r - f_{r-1}] =  f_{\hat{r}} - f_{\widehat{r+1}}. 
\end{equation}
The momentum $\hat{a}$ is generic, however if  
\begin{equation}
  \hat{a}\capstar 0 = k 
  \quad\Rightarrow\quad 
  \Deltastar_r(\hat{a},0) = \delta_{r,k}
\end{equation}
then 
\begin{equation}
  f(\hat{a}) |_{\hat{a}\capstar 0 = k} = \frac{1}{\sqrt{p_0}}\, f_{\hat{k}},
  \qquad k = 0, \dotsc, R+1.
\end{equation}
and the RFT is {\sl monochromatic}. 
The quantity 
\begin{equation}
f_{\hat{k}} = \sum_{r=k}^{R+1} p_r \bigl[f_r - f_{r-1}\bigr]
\end{equation}
defined through Eq. (\ref{eq:1rep}) is called the RFT 
of $f_r$. Similar to ordinary Fourier Transform, the wave number 
$\hat{a}\capstar 0 = k$ fixes the {\sl  resolution} of the RFT. Indeed
$\hat{a}\capstar 0 = k$ implies that $\hat{a}_t = 0$ for $t=k,\dotsc, R$, 
but $\hat{a}_{k-1} \not = 0$, and  shells $S_t$ with $t\geq k$ are averaged out.
By increasing the  value of $k$ we move down to lower shells and get more 
and more details.

The inverse RFT follows directly from Eq. (\ref{eq:1rep}), which together with
Eq. (\ref{eq:1repfun}) gives:
\begin{equation}
  f(a)  = \sum_{r=0}^{R+1}\, \frac{1}{p_r}[f_{\hat{r}} - f_{\widehat{r+1}}]\, 
             \delta_r(a,0)
        =  \sum_{r=0}^{R+1} f_{\hat{r}}\, \pbar_r(a,0).
\end{equation}
Moreover, iterating Eq. (\ref{eq:1rep}), leads to
\begin{equation}
 \label{eq:invRFT}
  f_r = \sum_{k=0}^r \frac{1}{p_k}[f_{\hat{k}} - f_{\widehat{k+1}}]
\end{equation}
which express $f_r$ in terms of its RFT $f_{\hat{k}}$.

\subsection{Two-replicas functions}
The case of functions depending on two replicas can be worked out in a 
similar way. 
Consider a function (matrix) $A^{ab} \equiv A(a,b)$ that depends 
on the overlap $r$ between the replicas $a$ and $b$:  
$A^{ab} = A_r$ if $a\cap b = r$. 
Then, as before [cfr. Eq. (\ref{eq:1repfun})]:
\begin{equation}
\label{eq:matrix}
 A^{ab} = \sum_{r=0}^{R+1} A_r\, \Delta_r(a,b)
              = \sum_{r=0}^{R+1} \bigl[A_r - A_{r-1}\bigr]\, \delta_r(a,b).
\end{equation}
The RFT of $A^{ab}$ can be performed in sequence, e.g., first on index 
$b$ and then on index $a$. The calculation is straightforward and leads 
to:
\begin{equation}
\label{eq:RFTmat}
  \begin{split}
  A(\hat{a},\hat{b}) &= \sum_{r=0}^{R+1} p_r \bigl[A_r - A_{r-1}\bigr]\, 
                        \deltastar_r(\hat{a},0)\, 
                        \delta_{R+1}(\hat{a}+\hat{b}, 0)\\
                    & = \sum_{r=0}^{R+1} A_{\hat{r}}\,\Deltastar_r(\hat{a},0)\,
                        \delta_{R+1}(\hat{a}+\hat{b}, 0),
   \end{split}
\end{equation}
where $A_{\hat{r}}$ is the RFT of $A_r$ introduced as before from the relation:
\begin{equation}
\label{eq:1idxRFT}
  p_r \bigl[A_r - A_{r-1}\bigr] = A_{\hat{r}} - A_{\widehat{r+1}}.
\end{equation}
As it happens for the  Fourier Transform of translation invariant functions, 
the dependence of $A^{ab}$ on the overlap, or co-distance, between replicas  leads
to the vanishing of the total momentum $\hat{a}+\hat{b}$ in the Replica Fourier Space
ensured by the delta function.\footnote{
Due to the periodicity of indexes $\delta_{R+1}(\hat{a}+\hat{b}, 0)$ means 
that $\hat{b}_r = -\hat{a}_r \pmod  {p_r/p_{r+1}}$.
If assume that $\hat{a}_r$ takes symmetric positive and negative
values, as wave-numbers in ordinary Discrete Fourier Transform, then  
$\delta_{R+1} (\hat{a}+\hat{b}, 0)= \delta_{\hat{a},-\hat{b}}$ 
with the usual Kroneker delta.} 

The RFT $A_{\hat{k}}$ is equal to the monocromatic $\hat{a}\capstar 0 = k$ 
RFT of $A^{ab}$:
\begin{equation}
\label{eq:hatA-k}
 A(\hat{a},-\hat{a})|_{\hat{a}\capstar 0 = k}  = A_{\hat{k}} 
                       = \sum_{r=k}^{R+1} p_r \bigl[A_r - A_{r-1}\bigr].
\end{equation}
In terms of the components $A_r$ the RFT of $A(a,b)$ can be also expressed as 
\begin{equation}
\label{eqLRFTmat1}
  A(\hat{a},\hat{b}) = \sum_{r=0}^{R+1} A_r\, \pstar_r(\hat{a},0)\,
                       \delta_{R+1}(\hat{a}+\hat{b}, 0).
\end{equation} 
The RFT can be easily inverted using Eq. (\ref{eq:1idxRFT}) 
obtaining: 
\begin{equation}
\label{eq:A-r}
  A_r = \sum_{k=0}^{r} \frac{1}{p_k} \bigl[A_{\hat{k}} - A_{\widehat{k+1}} 
                                     \bigr].
\end{equation}
and
\begin{equation}
\label{eq:matrix_1}
\begin{split}
 A^{ab}  &= \sum_{k=0}^{R+1} \frac{1}{p_k} 
            \bigl[A_{\hat{k}} - A_{\widehat{k+1}} \bigr]\,
            \delta_k(a,b)\\ 
         &=  \sum_{k=0}^{R+1}  A_{\hat{k}}\, \pbar_k(a,b).
 \end{split}
\end{equation}

The RFT shares many properties of the usual Fourier Transform. For example from 
Eq. (\ref{eq:inv}) it follows that the product of two matrices
\begin{equation}
\label{eq:convol0}
  C^{ab} = \sum_{c} A^{ac}\ B^{cb}
\end{equation}
can be written as
\begin{equation}
  C^{ab} = \sum_{c,c',\hat{c}} A^{ac} \chi(c\hat{c})\chi(-\hat{c}c')  B^{c'b}
               = \sum_{\hat{c}} A(a,\hat{c})\,B(-\hat{c},b)
\end{equation}
where $A(a,\hat{c})$ is the RFT of $A^{ac}$ on index $c$ and momentum 
$\hat{c}$, and similarly for $B$. 
Taking now the RFT on indexes $a$ and $b$ we recover the convolution relation
\begin{equation}
\label{eq:convol}
  C(\hat{a},\hat{b}) = \sum_{\hat{c}} A(\hat{a},\hat{c})\ B(-\hat{c},\hat{b}),
\end{equation}
from which it follows that
\begin{equation}
\label{eq:convolRFT}
 C_{\hat{k}} = A_{\hat{k}}\, B_{\hat{k}}.
\end{equation}
These relations are valid for any matrix $A^{ab}$ and $B^{ab}$. 
If $B = A^{-1}$ then $C^{ab} = \delta_{ab}$ and $\hat{C}_k = 1$, 
and the above relation gives
\begin{equation}
\label{eq:invmatRFT}
 A^{-1}_{\hat{k}} = 1/A_{\hat{k}}
\end{equation}
Finally  the inversion formula (\ref{eq:A-r}) leads to the following simple
expression
\begin{equation}
\label{eq:invmatrix}
  A^{-1}_r = \sum_{k=0}^{r} \frac{1}{p_k} 
         \left[\frac{1}{A_{\hat{k}}} - \frac{1}{A_{\widehat{k+1}}} \right],
\end{equation}
for the components of the inverse matrix $A^{-1}$.
In a similar way one can compute the components of the $m$-power of a matrix,
finding:
\begin{equation}
  (A^m)_r = \sum_{k=0}^{r} \frac{1}{p_k} 
          \left[\hat{A}_k^m - \hat{A}_{k+1}^m \right].
\end{equation}
 
The trace of a matrix can be computed either using the identity
\begin{equation}
  \sum_{a} A(a,a) = \sum_{\hat{a}} A(\hat{a},-\hat{a})
\end{equation}
or from Eq. (\ref{eq:matrix_1}) by noticing that  
$\pbar_k(a,a) = 1/p_k - 1/p_{k-1}$, or directly from 
the identity $\mbox{Tr}\, A = p_0 A_{R+1}$ using Eq. (\ref{eq:A-r}).
In all cases one ends up with 
\begin{equation}
\begin{split}
 \mbox{Tr}\, A  &= p_0 \sum_{k=0}^{R+1} \frac{1}{p_k} 
                     \bigl[ A_{\hat{k}} - A_{\widehat{k+1}}\bigr]
                         \\
              &= p_0 \sum_{k=0}^{R+1} 
              \left[\frac{1}{p_k} -\frac{1}{p_{k-1}}\right]\, A_{\hat{k}}
  \end{split}
\end{equation}
where the terms $1/p_{k-1}$ is missing for $k=0$.

\subsection{Three-replicas functions}
When the function depends on the overlap among more than two replicas 
different geometries are possible.
We shall consider the case of functions $A(a,b,c) \equiv A^r(a,b;c)$ 
in which the overlap $a\cap b=r$ between replicas $a$ and $b$ is held fixed,
and passive, while the cross-overlap with the replica $c$ can vary.
The ultrametric property of the metric implies that  $A^r(a,b;c)$ can depend on only  
one  cross-overlap:
\begin{equation}
  A^r(a,b;c) = A^r_t,
  \qquad t = \max(a\cap c, b\cap c),
\end{equation}
as shown in the tree representation of Fig. \ref{fig:3replica}.
By exploiting the symmetry of the function under the exchange of indexes $a$ and $b$, 
$A^r(a,b;c)$ is decomposed over the shells $S_t$ as:%, see Fig. \ref{fig:3replica}:
\begin{equation}
\begin{split}
 A^r(a,b;c) &=  \frac{1}{2}\sum_{t=0}^{r-1} A^r_t \bigl[\Delta_t(a,c) + \Delta_t(b,c)\bigr]
                         \,\Delta_r(a,b)\\
                    &%\phantom{=}
                    +  \frac{1}{2} A^r_r\bigl[\Delta_r(a,c)  - \delta_{r+1}(b,c)
                                                        +   \Delta_r(b,c)  - \delta_{r+1}(a,c)
                                                      \bigr]
                         \,\Delta_r(a,b)\\
                  & %\phantom{===}
                   +  \sum_{t=r+1}^{R+1} A^r_t \bigl[\Delta_t(a,c) + \Delta_t(b,c)\bigr]
                         \,\Delta_r(a,b)
\end{split}
\end{equation}
%
\if 0
\begin{equation}
\begin{split}
 A^r(a,b;c) &=  \frac{1}{2}\sum_{t=0}^{r} A^r_t \bigl[\Delta_t(a,c) + \Delta_t(b,c)\bigr]
                   -\frac{1}{2} A^r_r\bigl[\delta_{r+1}(a,c) + \delta_{r+1}(b,c)\bigr]
                  \\ 
                  & \phantom{===} +  \sum_{t=r+1}^{R+1} A^r_t \bigl[\Delta_t(a,c) + \Delta_t(b,c)\bigr]
\end{split}
\end{equation}
\fi
The symmetry factor  $1/2$ in the first and second terms accounts for the symmetry
of the tree for $t\leq r$ under the exchange of $a$ and $b$.
The second term enforces the constraint that the indexes
$a_r$, $b_r$ and $c_r$ are all different. Finally the last term is the contribution from the two
different configurations of the tree for $t>r$. 
By rearranging the terms and defining
\begin{equation}
\label{eq:pone}
  p_t^{(r)} = \left\{ \begin{array}{ll}
                                   p_t & t \leq r, \\
                                2 p_t & t > r,
                                \end{array}
                     \right.
\end{equation}
$A^r(a,b;c)$ can be written as
\begin{equation}
\label{eq:3rep}
  A^r(a,b;c) = \frac{1}{2}\sum_{t=0}^{R+1} \frac{p_t^{(r)}}{p_t} \bigl[ A^r_t - A^r_{t-1}\bigr] 
                                                \bigl[\delta_t(a,c) + \delta_t(b,c)\bigr]\, 
                                              \Delta_r(a,b).
\end{equation}

The RFT on index $c$  is easy evaluated using Eq. (\ref{eq:deltachi}) and reads:
% [cfr. eq. (\ref{eq:RFTmat})]
\begin{equation}
 A^r(a,b;\hat{c}) = \frac{1}{2} 
                                \sum_{t=0}^{R+1} A^r_{\hat{t}}\, \Deltastar_t(\hat{c},0)\,
                                \bigl[ \chi(a\hat{c}) + \chi(b\hat{c})\bigr]\,
                                \Delta_r(a,b)
\end{equation}
where 
\begin{equation}
  A^r_{\hat{t}} = \sum_{k=t}^{R+1} p_k^{(r)}\big[ A^r_k - A^r_{k-1}\bigr],
\end{equation}
defined through the relation
\begin{equation}
\label{eq:3repRFT}
  p_t^{(r)}\big[ A^r_t - A^r_{t-1}\bigr] = A^r_{\hat{t}} - A^r_{\widehat{t+1}}.
\end{equation}
is the RFT of $A^r_t$ with respect to the cross-overlap $t$ with passive overlap $r$.
%
\if 0
\begin{figure}[t]
\centering
\includegraphics[scale=0.85]{3momenta}
\protect\caption{Configurations of $A^r(\hat{a},\hat{b};\hat{c})$.
                 The thick line denotes zero momentum, while the 
                  horizontal tick the shell $\overset{*}S_r$ below which, but not including it,  the 
                  momentum must vanishes.
                  On each shell $\overset{*}S_{k=0,\dotsc,R}$
                  the momenta must satisfy the relation 
                   $\hat{a}_k + \hat{b}_k + \hat{c}_k = 0$.
                 }
\label{fig:3momenta}
\end{figure}
\fi
%
The RFT on indexes $a$ and $b$ is now evaluated using Eq. (\ref{eq:pstar_1}) and leads 
to the full RFT of $A^{r}(a,b;c)$:
\begin{equation}
\label{eq:3replRFT}
 A^r(\hat{a},\hat{b};\hat{c}) =                                        
                                       \frac{1}{2\sqrt{p_0}}\,
                                          \sum_{t=0}^{R+1} A^r_{\hat{t}}\, \Deltastar_t(\hat{c},0)\,
                                          \bigl[\pstar_r(\hat{a},0) + \pstar_r(\hat{b},0)\bigr]\,
                                          \delta_{R+1}(\hat{a}+\hat{b}+\hat{c},0).
\end{equation}
%
%The graphical tree representation of $A^r(\hat{a},\hat{b};\hat{c})$ is shown in 
%Fig. \ref{fig:3momenta}.

Inverting the RFT with the use of  Eq. (\ref{eq:3repRFT}),  leads to 
\begin{equation}
\begin{split}
  A^r(a,b;c) &= \frac{1}{2}\sum_{k=0}^{R+1} \frac{1}{p_k} 
                              \bigl[ A^r_{\hat{k}} - A^r_{\widehat{k+1}}\bigr] 
                                                \bigl[\delta_k(a,c) + \delta_k(b,c)\bigr]\, 
                       \Delta_r(a,b) \\                   
                      &= \frac{1}{2}\sum_{k=0}^{R+1} A^r_{\hat{k}}\, 
                         \bigl[  \pbar_k(a,c) +  \pbar_k(b,c)
                         \bigr]\,
                                              \Delta_r(a,b).
 \end{split}
\end{equation}
and
\begin{equation}
  A^r_{t} = \sum_{k=0}^{t} \frac{1}{p_k^{(r)}}\big[ A^r_{\hat{k}} - A^r_{\widehat{k+1}}\bigr] 
\end{equation}
which express $A^r(a,b;c)$, and $A^r_t$, in terms of $A^r_{\hat{k}}$.

\subsection{Four-replica functions}
\begin{figure}[t]
\centering
\includegraphics[scale=0.6]{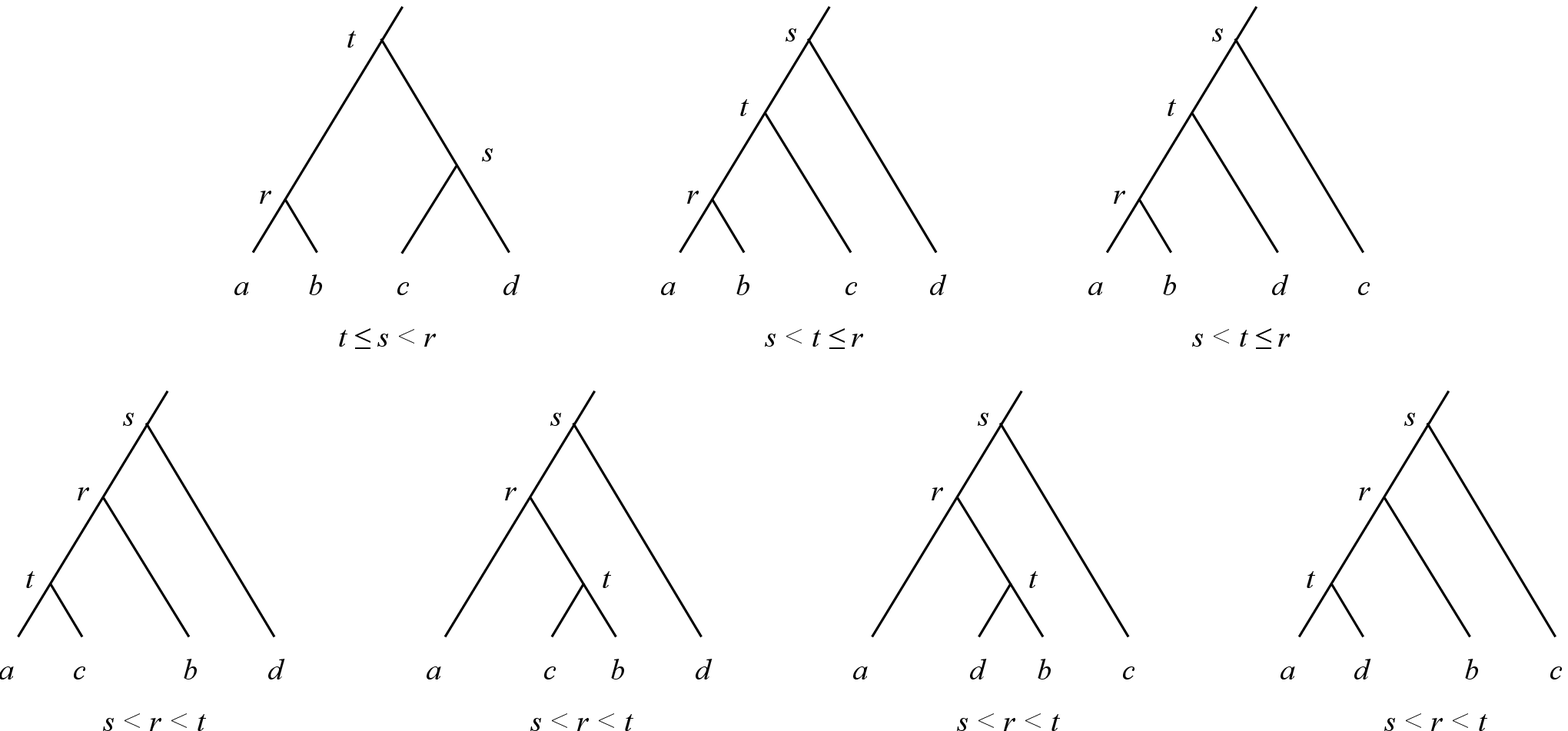}
\caption{Four-replica Longitudinal-Anomalous configurations with $a\cap b = r > c\cap d = s$.
                 }
\label{fig:4replica-rgs}
\end{figure}

As an example of functions depending on the overlap among four replicas we shall consider
the functions $A(a,b,c,d) = A^{r,s}(a,b;c,d)$ in wich the overlaps $a\cap b = r$ and $c\cap d = s$ 
are fixed and passive, while the cross-overlaps between replicas $a,b$ and replicas $c,d$ 
can vary. 
Functions of this type are relevant for the study of the fluctuations in the 
spin glass problem.
From the ultrametic property of the metric  it follows that if $r \not= s$ then 
$A^{r,s}(a,b;c,d)$ can depend on only one cross-overlap:
\begin{equation}
  A^{r,s}(a,b;c,d) = A^{r,s}_t,
  \qquad t = \max(a\cap c, a\cap d, b\cap c, b\cap d),
\end{equation}
as shown for $r>s$ in the tree representation of Fig. \ref{fig:4replica-rgs}. 
In the theory of spin glasses these configuration are called 
{\sl Longitudinal-Anomalous} (LA).
The decomposition of $A^{r,s}(a,b;c,d)$ on shells $S_t$ can be obtained
directly from the tree representation of Fig. \ref{fig:4replica-rgs}. 
Then  for $r>s$, and omitting the 
term $\Delta_r(a,b)\, \Delta_s(c,d)$ for simplicity of writing, it reads:
\begin{equation}
\begin{split}
 A^{r,s}(a,b;c,d) &=  
                      \frac{1}{4}\sum_{t=0}^{s} A^{r,s}_t 
                                \bigl[\Delta_t(a,c) + \Delta_t(b,c) + \Delta_t(b,d) + \Delta_t(a,d)\bigr]
     \\
   &\phantom{=}
                      -\frac{1}{4} A^{r,s}_s
               \bigl[\delta_{s+1}(a,c) + \delta_{s+1}(a,d) + \delta_{s+1}(b,c) + \delta_{s+1}(b,d)\bigr]
                      \\
   &\phantom{=}
                      + \frac{1}{2}\sum_{t=s+1}^{r} A^{r,s}_t 
                                \bigl[\Delta_t(a,c) + \Delta_t(a,d) + \Delta_t(b,c) + \Delta_t(b,d)\bigr]
     \\
   &\phantom{=}
                      -\frac{1}{2} A^{r,s}_r
                                \bigl[\delta_{r+1}(a,c) + \delta_{r+1}(a,d) + \delta_{r+1}(b,c) + \delta_{r+1}(b,d)\bigr]
                      \\
   &\phantom{=}
                      + \sum_{t=r+1}^{R+1} A^{r,s}_t 
                                \bigl[\Delta_t(a,c) + \Delta_t(a,d) + \Delta_t(b,c) + \Delta_t(b,d)\bigr].
\end{split}
\end{equation}

The first sum from $0$ to $s$ is the contribution of the first diagram in the first row of
Fig. \ref{fig:4replica-rgs}. The diagram is symmetric under the exchange 
$a\leftrightarrow b$ or  $c\leftrightarrow d$ leading to the symmetry factor $1/4$.
The second sum from $s+1$ to $r$ is the contribution of the second and third diagrams of the 
first row,
and $1/2$ is the symmetry factor of the  residual $a\leftrightarrow b$ symmetry.
The last sum is the contribution from the four diagrams in the second row. 
The  two terms with $\delta$'s ensure the correct geometry when $t=s,r$.
These can be understood as follows. When $t=s$ the indexes $c_s$ and $d_s$ must be 
different  from each other {\sl and} from $a_s$, first diagram of the first line. 
This leads to the term $\Delta_s(a,c) - \delta_{s+1}(a,d)$.
Symmetrizing for the exchange $a\leftrightarrow b$ or  $c\leftrightarrow d$
leads to the $t=s$ term in the first sum and to the residual term shown in the second line.
Similarly when $t=r$ 
the index $c_s$, second digram first row, or
the index $d_s$, third digram first row, 
must be different from $a_s$ {\sl and} $b_s$. Then we have a term
$\Delta_s(a,c) - \delta_{s+1}(b,c) + \Delta_s(a,d) - \delta_{s+1}(b,d)$. Symmetrization for 
the exchange $a\leftrightarrow b$ leads to the $t=r$ term of the second sum and
the residual term shown in third row.
 
Rearranging the terms,  introducing
\begin{equation}
\label{eq:p-rs}
  p_t^{(r,s)} = \left\{ \begin{array}{ll}
                                   p_t & t \leq s < r \\
                                2 p_t & s <  t \leq r \\
                                4 p_t & s <  r < t,
                                \end{array}
                     \right.
\end{equation}
and  the  $\Delta_r(a,b)\, \Delta_s(c,d)$ term,
the contribution to $A^{r,s}(a,b;c,d)$  from the LA configurations of 
Fig. \ref{fig:4replica-rgs} can be written in the compact from 
\begin{equation}
\label{eq:4rep-LA}
\begin{split}
  \leftidx{_{\rm LA}}A^{r,s}(a,b;c,d) = \frac{1}{4}\sum_{t=0}^{R+1} \frac{p_t^{(r,s)}}{p_t} 
                                      &\bigl[ A^{r,s}_t - A^{r,s}_{t-1}\bigr] 
                                                \bigl[\delta_t(a,c) + \delta_t(a,d) \\
                                           &     + \delta_t(b,c) + \delta_t(b,d)\bigr]\, 
                                              \Delta_r(a,b)\, \Delta_s(c,d).
\end{split}
\end{equation}
The subscript LA makes explicit that it follows from LA configurations.
This expression is valid also for $r<s$, provided the role of $r$ and $s$ 
in Eq. (\ref{eq:p-rs}) is exchanged. 
And, as it will be shown shortly, it remains  valid for $r= s$ too.
\begin{figure}[t]
\centering
\includegraphics[scale=0.6]{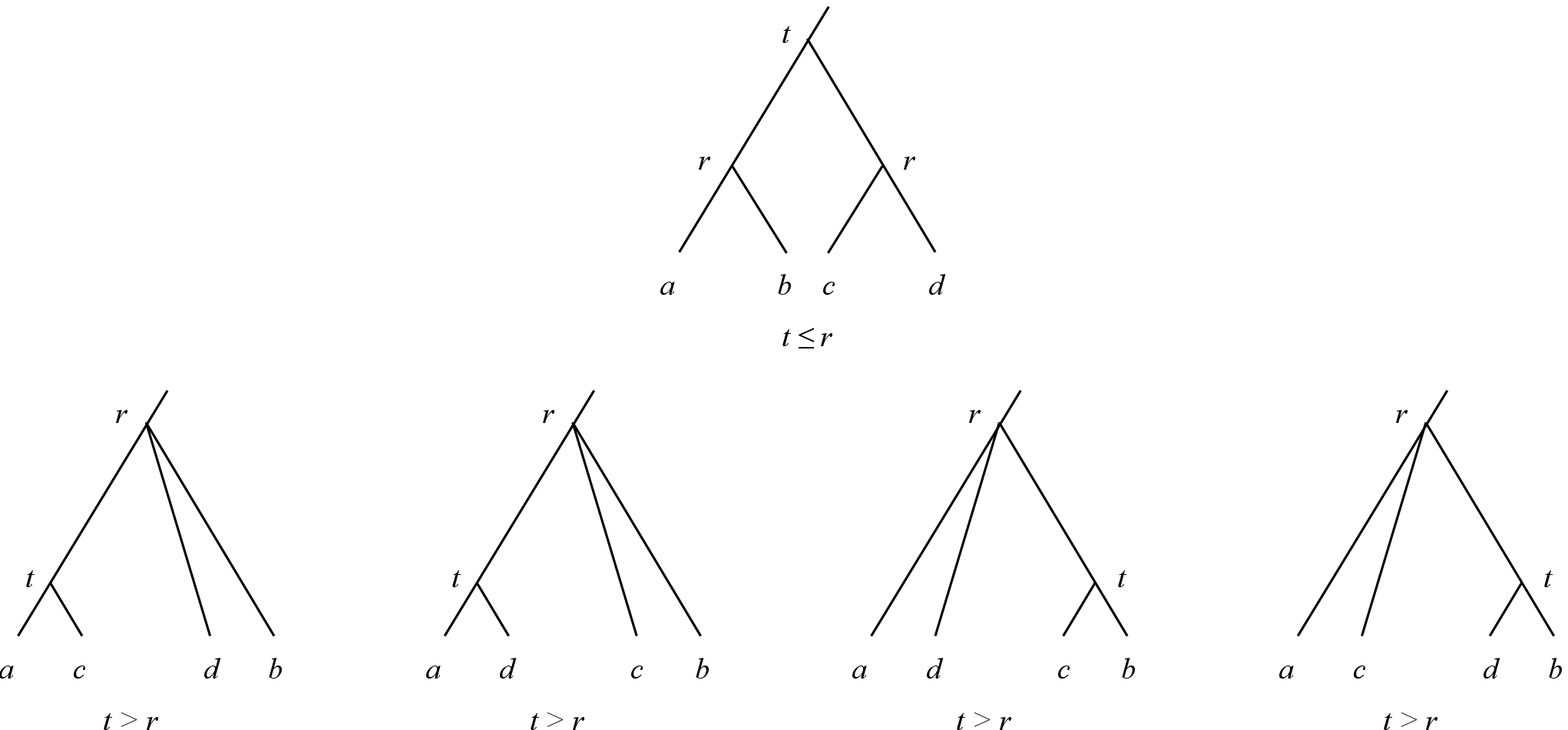}
\caption{Four-replica Longitudinal-Anomalous configurations with $a\cap b =  c\cap d = r$.
                 }
\label{fig:4replica-res}
\end{figure}

When $r=s$ different types of configurations are possible. 
The first type are   LA configurations shown in Fig. \ref{fig:4replica-res},
which depend on one cross-overlap:
\begin{equation}
  A^{r,r}(a,b;c,d) = A^{r,r}_t,
  \qquad t = \max(a\cap c, a\cap d, b\cap c, b\cap d).
\end{equation}
These configurations give:
\begin{equation}
\label{eq:4rep-res-LA}
\begin{split}
  \leftidx{_{\rm LA}}A^{r,r}(a,b;c,d) &=  
                      \frac{1}{4}\sum_{t=0}^{r-1} A^{r,r}_t 
                                \bigl[\Delta_t(a,c) + \Delta_t(a,d) + \Delta_t(b,c) + \Delta_t(b,d)\bigr]
     \\
   &%\phantom{=}
                      +\frac{1}{4} A^{r,r}_r
                    \bigl[\delta_{r}(a,c) + \delta_{r}(a,d) + \delta_{r}(b,c) + \delta_{r}(b,d)\bigr]
     \\
   &%\phantom{=}
                    - A^{r,r}_r
                    \bigl[\delta_{r+1}(a,c) + \delta_{r+1}(a,d) + \delta_{r+1}(b,c) + \delta_{r+1}(b,d)\bigr]
     \\
   &%\phantom{=}
                      + \sum_{t=r+1}^{R+1} A^{r,r}_t 
                                \bigl[\Delta_t(a,c) + \Delta_t(a,d) + \Delta_t(b,c) + \Delta_t(b,d)\bigr].
\end{split}
\end{equation}
The term $\Delta_r(a,b)\,\Delta_r(c,d)$ is again omitted for simplicity.
By rearranging the terms and defining
\begin{equation}
\label{eq:p-rr}
  p_t^{(r,r)} = \left\{ \begin{array}{ll}
                                   p_t & t \leq  r \\
                                4 p_t &  t  > r 
                                \end{array}
                     \right.
                    = \lim_{s\to r} p_t^{(r,s)} 
\end{equation}
this expression  reduces to the $r=s$ limit of Eq. (\ref{eq:4rep-LA}). (QED)

Note that in the configurations with $t > r$, second line Fig. \ref{fig:4replica-res} and fourth line
Eq. (\ref{eq:4rep-res-LA}), the cross-overlap between the two replicas not contributing to 
the cross-overlap $t$ is not restricted to $r$. For example, in the first diagram of the second line
of Fig. \ref{fig:4replica-res} configurations with $\delta_{r+1}(b,d)$ are allowed.\footnote{
This can be seen, e.g., by using the identity $\delta_r(b,d) = \Delta_r(b,d) + \delta_{r+1}(b,d)$.}
This leads us to the second type of configurations that appear for $r=s$
 when the cross-overlaps between $(a,b)$ and $(c,d)$ are larger than $r$.
These configurations require two cross-overlaps:
\begin{equation}
\begin{split}
  A^{r,r}(a,b;c,d) = A^{r,r}_{u,v},
   \qquad
        \begin{split}
                u &= \max(a\cap c, a\cap d),\\
                v &= \max(b\cap c, b\cap d),
         \end{split}
         \end{split}
\end{equation}
with $u,v \geq r+1$. Symmetry under the exchange of $a$ and $b$ implies 
$A^{r,r}_{u,v} = A^{r,r}_{v,u}$.

There are two contributions to $A^{r,r}_{u,v}$.  
The first one comes from the LA configurations in the second line of 
Fig. \ref{fig:4replica-res}  when the cross-overlap between the replicas
not contributing to the overlap $t$ is nor $r$. This reads
\begin{equation}
\label{eq:4rep-RLA}
  \leftidx{_{\rm LA}}A^{r,r}_{u,v} = A^{r,r}_u + A^{r,r}_v - A^{r,r}_r,
\end{equation}
where the last term corrects the double counting. By defining 
$A^{r,r}_{u,r} = A^{r,r}_{r,u} = A^{r,r}_u$ this term can be seen as a {\sl surface} or 
{\sl boundary} term.

\begin{figure}[t]
\centering
\includegraphics[scale=0.8]{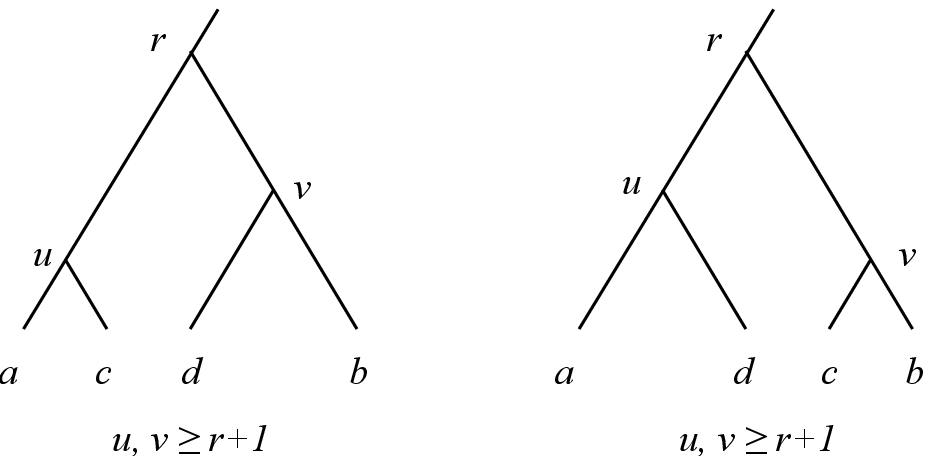}
\caption{Four-replica Replicon configurations with $a\cap b =  c\cap d = r$.
                 }
\label{fig:4replica-rep}
\end{figure}

The second type of configurations, 
called {\sl Replicon}\footnote{We use the full name to avoid confusion with 
$R$, the number of replica symmetry breaking.}  
in the theory of spin glasses,
 are those shown in Fig. \ref{fig:4replica-rep}.
The Replicon contribution reads: 
\begin{equation}
\label{eq:4rep-R}
\begin{split}
 &\leftidx{_{\rm R}}A^{r,r}(a,b;c,d) = \\
                                &\phantom{===}  =
                       \sum_{u=r+1}^{R+1}\sum_{v=r+1}^{R+1} 
                               \leftidx{_{\rm R}}A^{r,r}_{u,v} \,
                                \Bigl[\Delta_u(a,c)\,\Delta_v(b,d) 
%                                \\
 %                             &\phantom{============} 
                                +  \Delta_u(a,d)\,\Delta_v(b,c)\Bigr] \Delta_r(a,b)\,%\Delta_r(c,d)
                                \\
                                &\phantom{===} = 
  \sum_{u=r+1}^{R+1}\sum_{v=r+1}^{R+1} \llbracket A\rrbracket^{r,r}_{u,v}\, 
               \Bigl[\delta_u(a,c)\,\delta_v(b,d)
%                                \\
%                                &\phantom{============}                
                + \delta_u(a,d)\,\delta_v(b,c)\Bigr]\,
                    \Delta_r(a,b)\, %\Delta_r(c,d)
\end{split}
\end{equation}
where we have introduced the short-hand notation
\begin{equation}
  \llbracket A\rrbracket^{r,r}_{u,v} =  A^{r,r}_{u,v} - A^{r,r}_{u-1,v} - A^{r,r}_{u,v-1} + A^{r,r}_{u-1,v-1}.
\end{equation}
Symmetry under the exchange of $a$ and $b$ follows from $A^{r,r}_{u,v} = A^{r,r}_{v,u}$.

In the second equality there are no boundary terms because $\leftidx{_{\rm R}}A^{r,r}_{u,v}$ 
vanishes if $u$ or $v$ are equal to $r$, or smaller.
Note also that the second equality is written in terms of full $A^{r,r}_{u,v}$:
\begin{equation}
  A^{r,r}_{u,v} = \leftidx{_{\rm LA}}A^{r,r}_{u,v} + \leftidx{_{\rm R}}A^{r,r}_{u,v}.
\end{equation}
%and not of $ \leftidx{_{\rm R}}A^{r,r}_{u,v}$. 
The LA part depends only on one lower index at a time, 
then $\llbracket \leftidx{_{\rm LA}}A\rrbracket^{r,r}_{u,v} = 0$
and one  may use indifferently 
$\leftidx{_{\rm R}}A^{r,s}_{u,v}$ or $A^{r,s}_{u,v}$.

In conclusion $A^{r,s}(a,b;c,d)$  for generic $r$ and $s$  splits into the sum of
two terms:
\begin{equation}
\label{eq:4rep}
  A^{r,s}(a,b;c,d) =  \leftidx{_{\rm LA}}A^{r,s}(a,b;c,d) 
           + \leftidx{_{\rm R}}A^{r,r}(a,b;c,d)\, \delta_{r,s},
\end{equation}
with the LA and Replicon contributions  given by Eqs. (\ref{eq:4rep-LA}) and  (\ref{eq:4rep-R}), 
respectively, and hence
\begin{equation}
\label{eq:4repRFT}
  A^{r,s}(\hat{a},\hat{b};\hat{c},\hat{d}) =  \leftidx{_{\rm LA}}A^{r,s}(\hat{a},\hat{b};\hat{c},\hat{d}) 
           + \leftidx{_{\rm R}}A^{r,r}(\hat{a},\hat{b};\hat{c},\hat{d})\, \delta_{r,s}.
\end{equation}

The RFT of the LA part of $A^{r,s}(a,b;c,d)$ is equal to:
\begin{equation}
\label{eq:4repLARFT}
\begin{split}
 &\leftidx{_{\rm LA}}A^{r,s}(\hat{a},\hat{b};\hat{c},\hat{d}) =
        \frac{1}{4p_0}\,
        \sum_{t=0}^{R+1} A^{r,s}_{\hat{t}}\, \Deltastar_t(\hat{c}+\hat{d},0)  \\
    & \phantom{=}   
           \times
                                          \bigl[\pstar_r(\hat{a},0) + \pstar_r(\hat{b},0)\bigr]
                                          \bigl[\pstar_s(\hat{c},0) + \pstar_s(\hat{d},0)\bigr]\,
%    & \phantom{=}   
 %          \times
                 \delta_{R+1}(\hat{a}+\hat{b}+\hat{c}+\hat{d},0),
        \end{split}
\end{equation}
where  
\begin{equation}
\label{eq:LARFT}
  A^{r,s}_{\hat{k}} = \sum_{t=k}^{R+1} p_t^{(r,s)}\big[ A^{r,s}_t - A^{r,s}_{t-1}\bigr] 
\end{equation}
is the RFT of $A^{r,s}_t$ evaluated respect to the cross-overlap $t$ 
with passive direct overlaps  $r$ and $s$ given. 
Details can be found in \ref{app:4repLARFT}.
By using the identity:
\begin{equation}
\label{eq:4repRFTLA}
  p_t^{(r,s)}\big[ A^{r,s}_t - A^{r,s}_{t-1}\bigr] = A^{r,s}_{\hat{t}} - A^{r,s}_{\widehat{t+1}},
\end{equation}
the relation between $A^{r,s}_{\hat{t}}$ and $A^{r,s}_t$ can be inverted and one gets
the inverse RFT:
\begin{equation}
\label{eq:A-rs-t}
  A^{r,s}_{t} = \sum_{k=0}^{t} \frac{1}{p^{(r,s)}_k} 
          \bigl[A^{r,s}_{\hat{k}} - A^{r,s}_{\widehat{k+1}} \bigr].
\end{equation}
The RFT of the Replicon part of $A^{r,r}(a,b;c,d)$ reads: 
\begin{equation}
\label{eq:4repRRFT}
\begin{split}
 \leftidx{_{\rm R}}A^{r,r}(\hat{a},\hat{b};\hat{c},\hat{d}) &= 
       \frac{1}{p_0}\,
           \sum_{u=r+1}^{R+1}\sum_{v=r+1}^{R+1} 
                      A^{r,r}_{\hat{u},\hat{v}}\,
       \Deltastar_{u; r+1}(\hat{c},0)\,  \Deltastar_{v;r+1}(\hat{d},0) \\
    & \phantom{==}
     \times
                             \bigl[ \pstar_r(\hat{a}+\hat{c},0) + \pstar_r(\hat{b}+\hat{c},0) \bigr] \,
       \delta_{R+1}(\hat{a}+\hat{b}+\hat{c}+\hat{d},0),
\end{split}
\end{equation}
where
\begin{equation}
\label{eq:deltastar_1}
            \Deltastar_{k; r+1}(\hat{a},\hat{b})     = 
               \left\{
               \begin{aligned}
                 & \Deltastar_{k}(\hat{a},\hat{b}) & \qquad k > r+1, \\
                 & \deltastar_{r+1}(\hat{a},\hat{b})      & \qquad k = r+1.
               \end{aligned}
               \right.
\end{equation}
and 
\begin{equation}
\label{eq:dblRFT}
A^{r,r}_{\hat{k},\hat{l}} =  
  \sum_{u=k}^{R+1}\sum_{v=l}^{R+1} 
                  p_u\,p_v\,
                  \big[
                  A^{r,r}_{u,v} - A^{r,r}_{u-1,v} - A^{r,r}_{u,v-1} + A^{r,r}_{u-1,v-1}
                            \bigr],
%                            \qquad k,l \geq r+1.
\end{equation}
with $k,l \geq r+1$ 
is the {\sl double} RFT of $A^{r,r}_{u,v}$ on cross-overlaps  $u$, $v$ with passive overlap $r$.
Details can be found in \ref{app:4repRRFT}.
Again the double RFT can be evaluated indifferently with $A^{r,r}_{u,v}$ 
or $\leftidx{_{\rm R}}A^{r,r}_{u,v}$ since in  either case only the Replicon part contributes.
The quantities  $A^{r,r}_{u,v}$ and  $A^{r,r}_{\hat{u},\hat{v}}$ satisfy the relation:
\begin{equation}
\label{eq:dblRFT1}
\begin{split}
  p_u p_v \big[ A^{r,r}_{u,v} - A^{r,r}_{u-1,v} &- A^{r,r}_{u,v-1}  + A^{r,r}_{u-1,v-1} \bigr] \\
      & = 
  A^{r,r}_{\hat{u},\hat{v}} - A^{r,r}_{\widehat{u+1},\hat{v}} - A^{r,r}_{\hat{u},\widehat{v+1}}  
      + A^{r,r}_{\widehat{u+1},\widehat{v+1}},
\end{split}
\end{equation}
from which one easily gets the inverse RFT:  
\begin{equation}
\label{eq:invdblRFT}
\begin{split}
A^{r,r}_{u,v} - A^{r,r}_{u,r} &- A^{r,r}_{r,v} + A^{r,r}_{r,r} =
\leftidx{_{\rm R}}A^{r,r}_{u,v} 
\\
 &=   
  \sum_{k=r+1}^{u}\sum_{l=r+1}^{v} 
                  \frac{1}{p_k}\,\frac{1}{p_l}\,
                  \big[
                  A^{r,r}_{\hat{k},\hat{l}} - A^{r,r}_{\widehat{k+1},\hat{l}} 
                                       - A^{r,r}_{\hat{k},\widehat{l+1}} + A^{r,r}_{\widehat{k+1},\widehat{l+1}}
                            \bigr],
\end{split}                            
%                            \qquad u,v \geq r+1.
\end{equation}
expressing $\leftidx{_{\rm R}}A^{r,r}_{u,v}$,  in terms of its RFT  $A^{r,r}_{\hat{u},\hat{v}}$.

%\if 0
We note that with a bit of more work, one can show that
\begin{equation}
\label{eq:4rep-LA-P}
\begin{split}
  \leftidx{_{\rm LA}}A^{r,s}(a,b;c,d) = \frac{1}{4}\sum_{k=0}^{R+1}  
                                      A^{r,s}_{\hat{k}}
                                                &\bigl[\pbar_k(a,c) + \pbar_k(a,d) \\
                                           &     + \pbar_k(b,c) + \pbar_k(b,d)\bigr]\, 
                                              \Delta_r(a,b)\, \Delta_s(c,d).
\end{split}
\end{equation}
and
\begin{equation}
\label{eq:4repR1}
 \leftidx{_{\rm R}}A^{r,r}(a,b;c,d) = 
           \sum_{k,l=r+1}^{R+1}%\sum_{v=r+1}^{R+1} 
                      A^{r,r}_{\hat{k},\hat{l}}\,
                        \bigl[ \pbar_k(a,c)\,\pbar_l(b,d) + \pbar_k(b,c)\, \pbar_l(a,d)\bigr]\,
                  \Delta_r(a,b)
\end{equation}
where the term $(1/p_r)\,\delta_r(a,b)$ in the definition of $\pbar_k(a,b)$, Eq. (\ref{eq:Pr}), 
is missing if $k=r+1$.
These expressions can be taken as starting point to compute the eigenvalues of
$A^{r,s}(a,b;c,d)$  directly in the Replica Space following a procedure similar to that 
discussed in the next Section. This will essentially reproduce Ref. \cite{Temal94}. 
In the next Section we shall discuss the diagonalization in the RFT space. 
%\fi

\section{The Eigenvalue Problem}
\label{sec:EigenEq}

In this Section we illustrate the use of the RFT to solve eigenvalue equations. We shall consider 
only two cases: the two-replicas matrix $A^{ab}$, mainly to illustrate the procedure,  
and the four-replicas matrix $A^{ab,cd}$, relevant for the study of the stability of the static 
mean  field solution of  spin glass models.

\subsection{Two-replicas eigenvalue equation}
The eigenvalue equation for the matrix $A^{ab}$ is
\begin{equation}
\label{eq:eigeq2r}
  \sum_{b=1}^n A^{ab}\, \phi^b = \Lambda\, \phi^a,
  \qquad a = 1,\dotsc, n.
\end{equation}
where $A^{ab} = A_r$ if $a\cap b = r$ with $r=0,\dotsc, R+1$.
This case is rather simple because the RFT diagonalizes $A^{ab}$ as can be 
seen from, e.g.,  Eq. (\ref{eq:matrix_1}).
 However, in view of the more complex case of the  four-replicas matrix $A^{ab,cd}$, we shall
 discuss procedure of diagonalization in the Replica Space. 
 The eigenvalue equation in the Fourier Replica space reads
 \begin{equation}
  \sum_{\hat{b}} A(\hat{a},\hat{b})\, \phi(-\hat{b}) = \Lambda\, \phi(\hat{a}).
\end{equation}
where $A(\hat{a},\hat{b})$ is given by eq. (\ref{eq:RFTmat}).
To solve this equation we have to find a set of eigenvectors $\phi_k(\hat{a})$
which diagonalize the eigenvalue equation  for $\Lambda=\Lambda_k$. 
Consider the functions
\begin{equation}
\label{eq:psir}
  \psi_r (\hat{a};\hat{c}) = \Deltastar_r(\hat{c},0)\,\delta_{R+1}(\hat{a}+\hat{c},0),
\end{equation}
where $\hat{c}$ is a generic momentum  $\hat{c}\capstar 0 = r$, with 
$r$ ranging from $0$ to $R+1$.
These functions satisfies the identity
\begin{equation}
   \sum_{\hat{c}} \psi_r(\hat{a};\hat{c})\, \psi_s(\hat{b};\hat{-c}) =
     \delta_{r,s}\, \psi_r(\hat{a};\hat{b}),
\end{equation}
so that $A(\hat{a},\hat{b})$ can be written as
\begin{equation}
  A(\hat{a},\hat{b}) = \sum_{r=0}^{R+1} \sum_{\hat{c}}\, A_{\hat{r}}\, 
        \psi_r(\hat{a};\hat{c})\,\psi_r(\hat{b}; -\hat{c}).
\end{equation}
From this expression one readily sees that  
$\phi_k(\hat{a};\hat{e}) = \phi_{\hat{k}}\,\psi_k(\hat{a};\hat{e})$ with $\hat{e}\capstar 0 = k$, and 
$\phi_{\hat{k}}$ a constant, 
are the sought eigenvectors
with eigenvalues
\begin{equation}
  \Lambda_k = A_{\hat{k}}, 
  \qquad k = 0, \dotsc, R+1.
\end{equation}
For each $k$ the multiplicity of the eigenvalue $\Lambda_k$ is equal to the dimension of 
the subspace spanned by the eigenvectors $\phi_{k}(\hat{a};\hat{e})$, i.e.,  
the number of independent $\psi_{k}(\hat{a};\hat{e})$. 
From eq. (\ref{eq:psir}) this is equal to the number of $\hat{e}$ with $\hat{e}\capstar 0 = k$. 
By recalling that $\Deltastar_k(\hat{e},0)$ means that 
$\hat{e}_s$ is unrestricted for $s<k-1$ and equal to $0$ for $s\geq k$, and $\hat{e}_{k-1}\not= 0$,
the multiplicity $\mu(k)$ of $\Lambda_k$ is then
\begin{equation}
\label{eq:muk}
  \mu(k) = \frac{p_0}{p_1}\, \frac{p_1}{p_2}\cdots \frac{p_{k-2}}{p_{k-1}} 
                   \left(\frac{p_{k-1}}{p_k} - 1 \right)
               = 
               \left\{
               \begin{aligned}
                 &p_0  \left(\frac{1}{p_k} - \frac{1}{p_{k-1}} \right) & \qquad k > 0, \\
                 &\frac{p_0}{p_0} = 1                                                 & \qquad k = 0.
               \end{aligned}
               \right.
\end{equation}
Note that $\sum_{k=0}^{R+1}\mu(k) = p_0$, the total dimension $n$ of
the space.

In the Replica Space the eigenvectors read
\begin{equation}
\label{eq:eigenvect}
  \phi_k(a;e) = \sum_{\hat{a},\hat{e}} \phi_k(\hat{a})\, \chi(-\hat{a} a) \,\chi(-\hat{e} e)
                   = \phi_{\hat{k}}\, \pbar_k(a,e),
\end{equation}
with $e$ fixed [cfr. eq. (\ref{eq:matrix_1})], 
whose components on  shells $S_r$ are
\begin{equation}
 \phi_r = \sum_{s=0}^r \frac{1}{p_s} \bigl[\phi_{\hat{s}} - \phi_{\widehat{s+1}}\bigr]
       = \left\{
            \begin{aligned}
              0                                                                                           & \qquad r < k - 1,\\
              -\frac{1}{p_{k-1}}\, \phi_{\hat{k}}                                           & \qquad r = k -1, \\
              \left(\frac{1}{p_k} - \frac{1}{p_{k-1}}\right)\, \phi_{\hat{k}} & \qquad r > k -1.
            \end{aligned}
          \right.
\end{equation}
that are {\sl null} when the overlap $r$ is smaller than the resolution $k$ of the RFT, and
independent of $r$ when larger than $k$.

\subsection{Four-replicas eigenvalue equation: Hessian eigenvalues}
We now consider the eigenvalue equation for the matrix $A^{ab;cd}$. This problem is relevant for 
the stability analysis of spin glasses models, where $A^{ab;cd}$ is the Hessian
of the Gaussian fluctuations in the Replica Space about the saddle point  describing the static, 
thermodynamics, properties. We shall give an explicit example in the next Section.

The eigenvalue equation for $A^{ab;cd}$ is 
\begin{equation}
  \sum_{c\leq d}\, A^{ab;cd}\, \phi^{cd} = \Lambda\, \phi^{ab}
\end{equation}
where $\phi^{ab}$ is a {\sl vector} in a space of dimension $p_0\,(p_0+1)/2$.
By exploiting the symmetry under indexes exchange $c\leftrightarrow d$, 
the restriction on the sum is removed and  the eigenvalue equation becomes\footnote{In the 
Spin Glass problem diagonal terms are missing, so $r=R+1$ value is excluded and the space 
dimension of $\phi^{ab}$ is reduced from $p_0(p_0+1)/2$ to $p_0(p_0-1)/2$.}
 \begin{equation}
 \label{eq:eigeneq}
    \frac{1}{2}\sum_{cd} A^{ab;cd}\, \phi^{cd} = \Lambda\, \phi^{ab},
    \qquad a\cap b = r = 0,\dotsc, R+1,
 \end{equation}
with the implicit assumption that the diagonal term $A^{ab;cc}$ is multiplied by $2$ 
to compensate for the factor $1/2$.

This equation has been solved in the Replica Space by De Dominicis, Kondor and Temesv\`ari,  in 
Ref. \cite{Temal94}. The direct construction of the eigenvectors in the Replica Space 
is rather involved. 
Here, following the same procedure used for the two-replica matrix, we show that the construction 
of the eigenvectors in the Fourier Replica Space is straightforward.  
Moreover, the distinction between the LA  and the Replicon sectors  arises naturally.

To solve the eigenvalue equation in the  Replica Fourier Space,  we first rewrite the sum 
over $c$ and $d$ as
\begin{equation}
 \sum_{cd} \to \sum_{s=0}^{R+1} \sum_{cd | c\cap d = s}
\end{equation}
so that $A^{ab;cd} \to A^{r,s}(a,b;c,d)$, and then perform the RFT on the four indexes. 
The resulting eigenvalue equation is:
\if 0
 \begin{equation}
 \frac{1}{2} \sum_{s=0}^{R+1} \sum_{cd} A^{r,s}(a,b;c,d)\,\phi^{s}(c,d) = \Lambda\, \phi^r(a,b).
 \end{equation}
and then In term of RFT this equation reads
\fi
 \begin{equation}
 \label{eq:eigL1}
 \frac{1}{2} \sum_{s=0}^{R+1} 
      \sum_{\hat{c}\, \hat{d}} A^{r,s}(\hat{a},\hat{b}; \hat{c},\hat{d})\,
             \phi^{s}(-\hat{c}, -\hat{d}) = \Lambda\, \phi^r(\hat{a},\hat{b}).
 \end{equation}
where $A^{r,s}(\hat{a},\hat{b}; \hat{c}, \hat{d})$ given by Eqs. (\ref{eq:4repRFT}), 
(\ref{eq:4repLARFT}) and (\ref{eq:4repRRFT}).

As done for the case of the two-replicas matrix, to solve this equation one needs 
a set of independent functions $\psi^r(\hat{a},\hat{b})$ in the
$p_0(p_0+1)/2$ dimensional Fourier Replica Space in term of which express the eigenvectors.
The functions $\psi^r(a,b)$ in the Replica space must depend on the overlap $a\cap b = r$, 
an hence the total momentum in the Replica Fourier Space is conserved and equal to zero,
while $\psi^r(\hat{a},\hat{b})$  for different $r$ are orthogonal
\begin{equation}
\label{eq:ortho}
  \sum_{\hat{a},\hat{b}} \psi^r(\hat{a},\hat{b})\,\psi^s(\hat{a},\hat{b}) = 0 
  \quad\mbox{if}\ r\not=s.
\end{equation}

Inspection of  Eq. (\ref{eq:4repLARFT}) shows that by introducing the functions
\begin{equation}
\label{eq:psik}
   \psi^{r}_k(\hat{a},\hat{b};\hat{e}) = \frac{1}{2\sqrt{p_0}}
               \bigl[\pstar_r(\hat{a},0) + \pstar_r(\hat{b},0)\bigr]\,
               \Deltastar_k(\hat{e},0)\,
   \delta_{R+1}(\hat{a}+\hat{b}+\hat{e},0)
\end{equation}
with $r,k = 0, \dotsc, R+1$, then the LA part can be written in a form similar to that found for the
two-replicas matrix, i.e.:
\begin{equation}
   \leftidx{_{\rm LA}}A^{r,s}(\hat{a},\hat{b};\hat{c},\hat{d}) = 
        \sum_{t=0}^{R+1} A^{r,s}_{\hat{t}}\, 
        \overline{\psi}^{\,r,s}_t(\hat{a},\hat{b};\hat{c},\hat{d}),
\end{equation}
where 
\begin{equation}
\label{eq:bar-psi-rsk}
  \overline{\psi}^{\, r,s}_{k}(\hat{a},\hat{b};\hat{c},\hat{d}) =  \sum_{\hat{e}}
        \psi^r_k(\hat{a},\hat{b};\hat{e})\, \psi^s_k(\hat{c},\hat{d};-\hat{e}).
\end{equation}
The functions $\psi^{r}_k(\hat{a},\hat{b};\hat{e})$ satisfy the condition (\ref{eq:ortho}) and, 
moreover, functions with different $k$ are orthogonal:
%% Problem aux file
\begin{equation}
\label{eq:psik-psirl}
  \sum_{\hat{a},\hat{b}}  \psi^{s}_t(\hat{a},\hat{b};\hat{e})\,  
                                         \psi^{r}_k(-\hat{a},-\hat{b};-\hat{f})
                                         = \frac{1}{2} \delta_{r}^{(k-1)}\,
                                             \delta_{s,r}\, \delta_{t,k}\, \Deltastar_k(\hat{e},0)\,
                                             \delta_{R+1}(\hat{e},\hat{f}),
\end{equation}
where, for $\hat{e}\capstar 0 = k$, 
\begin{equation}
  \delta_{r}^{(k-1)} = \pstar_r(0,0) + \pstar_r(\hat{e},0) = p_r^{(k-1)} - p_{r+1}^{(k-1)}.
\end{equation}
Because of the constraint on the total momentum, and the symmetry of $\psi^r(\hat{a},\hat{b})$,
the number of linearly independent  functions $\psi^r_k(\hat{a},\hat{b};\hat{e})$ 
for each $r$ and $k$  is equal to  the number $\mu(k)$, Eq. (\ref{eq:muk}),
of $\hat{e}:  \hat{e}\capstar 0 = k$.
The dimension of the subspace spanned by these functions for  each $r$ is then
\begin{equation}
\label{eq:dimLAr}
 \mbox{Dim}_r[\psi^r_k] =  \sum_{k=0}^{R+1}\mu(k) = p_0, 
  \qquad r = 0,\dotsc, R+1,
\end{equation}
which gives a total dimension of $p_0 (R+2)$.

The Replicon part, Eq. (\ref{eq:4repRRFT}), can be written as
\begin{equation}
   \leftidx{_{\rm R}}A^{r,r}(\hat{a},\hat{b};\hat{c},\hat{d}) =   
	     \sum_{u=r+1}^{R+1} \sum_{v=r+1}^{R+1} A^{r,r}_{\hat{u},\hat{v}}\,
           \varphi^{r,r}_{u,v}(\hat{a},\hat{b};\hat{c},\hat{d})
\end{equation}
where
\begin{equation}
\label{eq:varphikl}
\begin{split}
 \varphi^{r,r}_{k,l}(\hat{a}, \hat{b}; \hat{c}, \hat{d}) &= \frac{1}{p_0}\bigl[
                        \pstar_r(\hat{a}+\hat{c},0) +  \pstar_r(\hat{b}+\hat{c},0)
                                                 \bigr]\,
                         \Deltastar_{k; r+1}(\hat{c},0)\, 
                         \Deltastar_{l;r+1}(\hat{d},0)\\
                         &\phantom{=========}
                         \times
                         \delta_{R+1}(\hat{a}+\hat{b} + \hat{c} +\hat{d}, 0)
   \end{split}
\end{equation}
with $k,l \geq r+1$ and $\Deltastar_{k; r+1}(\hat{a},\hat{b})$ defined in Eq. (\ref{eq:deltastar_1}).
These functions are not 
orthogonal to $\psi^r_k$ and mix with them when either one or both indexes 
$k$ and $l$ are equal to $r+1$:
\begin{equation}
\label{eq:Qv1}
  \sum_{\hat{c},\hat{d}}  \varphi^{r,r}_{u,v}(\hat{a},\hat{b};\hat{c}, \hat{d})\,  
                                         \psi^{s}_k(-\hat{c},-\hat{d};\hat{e})
                                         = \delta_{r,s}\, \delta_{(u,v)(r+1,m)}\, 
                                             \psi^{r}_k(\hat{a},\hat{b};\hat{e}),
\end{equation}
where $m = \max(r+1, k)$ and 
$\delta_{(a,b)(c,d)} = \delta_{a,c}\,\delta_{b,d} + \delta_{a,d}\,\delta_{b,c}$
is the ``symmetric'' delta function, and
\begin{equation}
\label{eq:Qv11}
  \sum_{\hat{a},\hat{b}}  \varphi^{r,r}_{u,v}(\hat{a},\hat{b};\hat{c}, \hat{d})\,  
                                         \psi^{s}_k(-\hat{a},-\hat{b};\hat{e})
                                         = 2\, \delta_{r,s}\, 
                                         \Deltastar_{u;r+1}(\hat{c},0)\, \Deltastar_{v;r+1}(\hat{d},0)\, 
                                             \psi^{r}_k(\hat{c},\hat{d};\hat{e}),
\end{equation}
Details can be found in \ref{app:Qv1}.
We then define  
\begin{equation}
\label{eq:psikl}
  \psi^{r,r}_{k,l}(\hat{a}, \hat{b}; \hat{c}, \hat{d}) = \varphi^{r,r}_{k,l}(\hat{a}, \hat{b}; \hat{c}, \hat{d}) 
   - \sum_{t=0}^{R+1}\frac{2}{\delta_r^{(t-1)}}\, \delta_{(k,l)(r+1,m)}\,
     \overline{\psi}^{\,r,r}_t(\hat{a},\hat{b}; \hat{c},\hat{d})
\end{equation}
where $m = \max(r+1, t)$,
so that 
\begin{equation}
\label{eq:psikl-psik}
  \sum_{\hat{c},\hat{d}}  \psi^{r,r}_{u,v}(\hat{a},\hat{b};\hat{c}, \hat{d})\,  
                                         \psi^{s}_k(-\hat{c},-\hat{d};\hat{e})
                                         = 0,
\end{equation}
as can be easily verified by using the identity: 
\begin{equation}
 \sum_{\hat{c},\hat{d}} \overline{\psi}^{\,r,s}_{t}(\hat{a},\hat{b};\hat{c},\hat{d})\,
                                        \psi^{l}_{k}(-\hat{c},-\hat{d};\hat{e}) =
                                        \frac{1}{2} \delta_{s,l}\,\delta_{t,k}\,
                                        \delta^{(k-1)}_{l}\,\psi^{r}_{k}(\hat{a},\hat{b};\hat{e}).
\end{equation}

To count the number of linearly independent $\psi^{r,r}_{k,l}$ we consider first the case $k,l>r+1$,
and  $\psi^{r,r}_{k,l} = \varphi^{r,r}_{k,l}$. Above level $r$ there are $p_0/p_r$ different ways of
choosing $\hat{a}$ and $\hat{b}$ with the constraint 
$\hat{a}_t + \hat{b}_t  + \hat{c}_t  +\hat{d}_t  = 0$. 
At level $r$, where we have the maximal resolution, the number of independent choices is  
equal to number of independent boxes $(\hat{a}_r, \hat{b}_r )$ that can be singled out
with $\hat{c}_r$ and $\hat{d}_r$.
This, for say fixed $\hat{c}_r$, leaves $(p_{r}/p_{r+1}-1)$ choiches for $\hat{d}_r$.  Thus the total 
number is equal to $\frac{1}{2}(p_r/p_{r+1})(p_r/p_{r+1} -1)$, because of the 
$(\hat{c}_r,\hat{d}_r)$ symmetry.  Below level $r$ the momenta satisfy the stronger 
constraints
$\hat{a}_t+\hat{c}_t=0$ and  $\hat{b}_t+\hat{d}_t=0$, or the symmetric one. 
The total number of independent choices is then equal to  
the number of $[\hat{c}_{r+1},\dotsc,\hat{c}_{k-1}']$, i.e.:
\begin{equation}
\begin{split}
 \#[\hat{c}_{r+1},\dotsc,\hat{c}_{k-1}'] &= 
 \frac{p_{r+1}}{p_{r+2}}\dotsc\frac{p_{k-2}}{p_{k-1}}
               \left( \frac{p_{k-1}}{p_k} -1 \right)\\ 
              & = p_{r+1}\, \left(\frac{1}{p_k} - \frac{1}{p_{k-1}} \right).
\end{split}
 \end{equation}
times the $\#[\hat{d}_{r+1},\dotsc,\hat{d}_{l-1}']$. Taken all together  these lead to 
the  multiplicity  
\begin{equation}
\label{eq:mukl}
 \mu(k,l) = \frac{1}{2} p_0 \bigl(p_r - p_{r+1}\bigr)\,
                                    \left(\frac{1}{p_k} - \frac{1}{p_{k-1}} \right)
                                    \left(\frac{1}{p_l} - \frac{1}{p_{l-1}} \right),
\end{equation}
for $k,l > r+1$.

When $k>r+1$ but $l=r+1$, then 
\begin{equation}
\label{eq:psikrp1}
\begin{split}
  &\psi^{r,r}_{k,r+1}(\hat{a}, \hat{b}; \hat{c}, \hat{d}) = 
  \frac{1}{p_0}\left\llbracket  \Bigl[
                        \pstar_r(\hat{a}+\hat{c},0) +  \pstar_r(\hat{b}+\hat{c},0)
                                                 \Bigr]\,
                        \Deltastar_{k}(\hat{c},0)\,  \deltastar_{r+1}(\hat{d},0)
                      \right.
\\
  &   \phantom{==}
                      \left.
   -\frac{1}{2(p_r - p_{r+1})} 
               \Bigl[\pstar_r(\hat{a},0) + \pstar_r(\hat{b},0)\Bigr]\, 
               \Bigl[\pstar_r(\hat{c},0) + \pstar_r(\hat{c},0)\Bigr]\, 
                   \Deltastar_{k}(\hat{c}+\hat{d},0)
                         \right\rrbracket\,                 
                                     \\
                                        &\phantom{==============}
               \times         \delta_{R+1}(\hat{a}+\hat{b} + \hat{c} +\hat{d}, 0).
\end{split}   
\end{equation}
The number of independent choices above level $r$ remains unaffected by the correction.
The second term, however, cancels one possible choice for $\hat{d}_r$ [for fixed $\hat{c}_r$], 
and the number of choices reduces to 
$(p_r/p_{r+1} -2)$. The total number then becomes 
$\frac{1}{2}(p_r/p_{r+1})(p_r/p_{r+1} -2)$.  
Below $r$ one momentum is identically equal to $\hat{d}_t=0$ while the other is fixed as before by 
$\hat{c}\capstar 0 = k$.
The multiplicity is then:
\begin{equation}
\label{eq:mukrp1}
 \mu(k>r+1,l=r+1) = \frac{1}{2} p_0 \bigl(p_r - 2\, p_{r+1}\bigr)\,
                                    \left(\frac{1}{p_k} - \frac{1}{p_{k-1}} \right)
                                    \frac{1}{p_{r+1}}.
\end{equation}
The case $k=r+1$ and $l > r+1$ is obtained by exchanging $k$ and $l$.

Finally for boundary case $k=l=r+1$, $\psi^{r,r}_{k,l}$ is equal to
\begin{equation}
\label{eq:psirp1rp1}
\begin{split}
  \psi^{r,r}_{r+1,r+1}&(\hat{a}, \hat{b}; \hat{c}, \hat{d}) = 
                \frac{1}{p_0}\left\llbracket  
                 \Bigl[ \pstar_r(\hat{a}+\hat{c},0) +  \pstar_r(\hat{b}+\hat{c},0) \Bigr]\,
                        \deltastar_{r+1}(\hat{c},0)\, \deltastar_{r+1}(\hat{d},0)
                        \right. \\
  &   \phantom{===}
        -  \Bigl[\pstar_r(\hat{a},0) + \pstar_r(\hat{b},0)\Bigr]\, 
           \Bigl[\pstar_r(\hat{c},0) + \pstar_r(\hat{d},0)\Bigr]
           \\
   & \phantom{=}
   \times
   \left.
   \left[        
              \frac{1}{p_r -2  p_{r+1}} 
               \Deltastar_{r+1}(\hat{c}+\hat{d},0)
               +
              \frac{1}{2(p_r - p_{r+1})} 
               \deltastar_{r}(\hat{c}+\hat{d},0)               
               \right]
                       \right\rrbracket
               \\
                                        &\phantom{============} \times
                        \delta_{R+1}(\hat{a}+\hat{b} + \hat{c} +\hat{d}, 0).
\end{split}   
\end{equation}
The two additional terms  give two constraints on level $r$, reducing
the choices to $\frac{1}{2}(p_r/p_{r+1})(p_r/p_{r+1} -3)$.  For  levels above $r$  
the number of choices is unchanged, while below $r$ both momenta have identically 
zero components.  The 
multiplicity  then reads:
\begin{equation}
\label{eq:murp1rp1}
 \mu(k=r+1,l=r+1) = \frac{1}{2} p_0 \bigl(p_r - 3\, p_{r+1}\bigr)\,
                                    \frac{1}{p_{r+1}^2}.
\end{equation}

The subspace spanned by the functions 
$\psi^{r,r}_{k.l}$ for fixed $r=0, \dotsc, R$ has then dimension:
\begin{equation}
 \mbox{Dim}_r[\psi^{r,r}_{k,l}] = \sum_{k=r+1}^{R+1}\sum_{l=r+1}^{R+1}\mu(k,l)
                                                 = \frac{p_0}{2} (p_{r} - p_{r+1} - 2).
\end{equation}
The subspace spanned by the functions $(\psi^r_k, \psi^{r,r}_{kl})$ for fixed $r$ 
has hence dimension
\begin{equation}
  \mbox{Dim}_r[\psi^{r}_{k}]  +  \mbox{Dim}_r[\psi^{r,r}_{k,l}]  =
  \left\{ \begin{array}{ll}
                                   \frac{1}{2}\, p_0\, (p_{r} - p_{r+1}), & r = 0,\dotsc R, \\                                   
                                  p_0, & r =R+1,
                                \end{array}
                     \right.
\end{equation}
and equals the dimension of the shell $S_r$, cfr. Eq. (\ref{eq:sumDelta}). The factor $1/2$ follows 
from the $a\leftrightarrow b$ symmetry.  
The functions $(\psi^r_k, \psi^{r,r}_{k,l})$ form then a complete basis and,
in terms of  them,  $A^{r,s}(\hat{a},\hat{b}; \hat{c},\hat{d})$ takes the form 
\begin{equation}
\label{eq:4rep-lin}
\begin{split}
  A^{r,s}(\hat{a},\hat{b}; \hat{c},\hat{d}) &= 
    \sum_{t=0}^{R+1} \left[
            \frac{4}{\delta^{(t-1)}_r}\, A^{r,r}_{\widehat{r+1},\hat{m}}\, \delta_{r,s} + A^{r,s}_{\hat{t}} 
                                                              \right]\,
                                     \overline{\psi}^{\, r,s}_{t}(\hat{a},\hat{b};\hat{c},\hat{d}) \\
      &\phantom{===}
      +  \delta_{r,s}\sum_{u=r+1}^{R+1} \sum_{v=r+1}^{R+1}\, 
         A^{r,r}_{\hat{u},\hat{v}}\, 
      \psi^{r,r}_{u,v} (\hat{a},\hat{b}; \hat{c},\hat{d}),
\end{split}                                                              
\end{equation}
where $m = \max(r+1, t)$. 
This form is valid for any $r$ and $s$, provided the Replicon 
contribution $A^{r,r}_{\hat{u},\hat{v}}$ is taken equal to zero for $r=R+1$. 

By introducing the function:
\begin{equation}
\label{eq:s-psikl}
\begin{split}
           \overline{\psi}^{\, r,r}_{u,v}(\hat{a},\hat{b};\hat{c},\hat{d}) &= \frac{1}{2}
           \bigl[
            \psi^{r,r}_{u,v}(\hat{a},\hat{b};\hat{c},\hat{d}) +
             \psi^{r,r}_{v,u}(\hat{a},\hat{b};\hat{c},\hat{d})
           \bigr]
           \\
   &= \frac{1}{2}
           \bigl[
            \psi^{r,r}_{u,v}(\hat{a},\hat{b};\hat{c},\hat{d}) +
             \psi^{r,r}_{u,v}(\hat{a},\hat{b};\hat{d},\hat{c})
           \bigr],        
\end{split}           
\end{equation}
symmetrical in both the first and second pair of indexes, and using the identity:
\begin{equation}
\label{eq:s-psikl-s-psikl}
  \sum_{\hat{e},\hat{f}}  \overline{\psi}^{\, r,r}_{u,v}(\hat{a},\hat{b};\hat{e}, \hat{f})\,  
                                         \overline{\psi}^{\, s,s}_{k,l}(\hat{c},\hat{d};-\hat{e}, -\hat{f})
                                         = \delta_{r,s}\, \delta_{(u,v)(k,l)}\,
                                         \overline{\psi}^{\, r,r}_{k,l}(\hat{a},\hat{b};\hat{c}, \hat{d}).                
\end{equation}
$A^{r,s}(\hat{a},\hat{b}; \hat{c},\hat{d})$ then takes the equivalent quadratic form:
\begin{equation}
\label{eq:4rep-quad}
\begin{split}
  A^{r,s}(\hat{a},\hat{b}; \hat{c},\hat{d}) &= 
    \sum_{t=0}^{R+1}\sum_{\hat{e}} \left[
            \frac{4}{\delta^{(t-1)}_r}\, A^{r,r}_{\widehat{r+1},\hat{m}}\, \delta_{r,s} + A^{r,s}_{\hat{t}} 
                                                              \right]\,
                                                              \psi^{r}_{t}(\hat{a},\hat{b};\hat{e})\,
                                                              \psi^{s}_{t}(\hat{c},\hat{d};-\hat{e}) \\
      &
      +  \delta_{r,s}\sum_{u,v=r+1}^{R+1}\, 
         \sum_{\hat{e},\hat{f}} A^{r,r}_{\hat{u},\hat{v}}\, 
      \overline{\psi}^{\, r,r}_{u,v} (\hat{a},\hat{b}; \hat{e},\hat{f})\,                                                             
      \overline{\psi}^{\, r,r}_{u,v} (-\hat{e},-\hat{f}; \hat{c},\hat{d}).                                                           
\end{split}                                                              
\end{equation}

The introduction of the functions $(\psi^r_k, \psi^{r,r}_{k,l})$ splits
the eigenvalue equation (\ref{eq:eigL1}) into two independent parts, called
the LA and Replicon sectors, associated with $\psi^r_k$ and $\psi^{r,r}_{k,l}$, respectively.
The matrix $A^{ab,cd}$ can then be diagonalized independently in each sector.

\bigskip
\noindent
$\bullet$ {\sl The LA Sector} \\
The eigenvectors in the LA sector  are
 $\phi^{r}_k(\hat{a},\hat{b}) =  \phi^r_{\hat{k}} \psi^{r}_k(\hat{a},\hat{b};\hat{e})$ and, 
using  Eqs. (\ref{eq:psik-psirl}), (\ref{eq:psikl-psik}) and  the identity 
\begin{equation}
 \frac{1}{2}  \sum_{\hat{c}\, \hat{d}} A^{r,s}(\hat{a},\hat{b}; \hat{c},\hat{d})\,
                                  \psi^{s}_{t}(-\hat{c},-\hat{d};\hat{e}) =
                     \left[
                A^{r,r}_{\widehat{r+1},\hat{m}}\, \delta_{r,s} + \frac{1}{4}\delta^{(t-1)}_r\,A^{r,s}_{\hat{t}} 
                                                              \right]\,
                                                              \psi^{r}_{t}(\hat{a},\hat{b};\hat{e}),             
\end{equation}
the eigenvalue equation becomes:
\begin{equation}
\label{eq:eigenLA}
 \sum_{s=0}^{R+1}
   \Bigl[ 
   \delta_{r,s} A^{r,r}_{\widehat{r+1},\hat{m}} + \frac{1}{4}\delta^{(k-1)}_s A^{r,s}_{\hat{k}}
   \Bigr]\, \phi^s_{\hat{k}} 
   = \Lambda\, \phi^{r}_{\hat{k}},
\end{equation}
where $m = \max(k, r+1)$ and  $r,k = 0,\dotsc, R+1$. 
The case $k=0$ identifies the Longitudinal sector, for which the eigenvalue equation in the 
Replica Space assumes the form given in Eq. (\ref{eq:eigeneq}). For the Anomalous
sector $k>0$ the eigenvalue equation in the Replica Space reds instead:
\begin{equation}
    \frac{1}{2}\sum_{cd} A^{ab;cd}\,\phi^{cd;e} = \Lambda\, \phi^{ab;e},
    \qquad a\cap b = r = 0,\dotsc, R+1,
\end{equation}
where $e$ is a fixed passive replica.

In the LA sector the eigenvalue equation is not diagonalized by the RFT, but rather reduced to
a set of $R+2$ independent $(R+2)\times(R+2)$ blocks labelled by $k$
({\sl block} diagonalization). 
For each $k=0,\dotsc, R+1$ there are $R+2$ eigenvalues with the multiplicity:
\begin{equation}
  \mu(k)  = 
               \left\{
               \begin{aligned}
                 &p_0  \left(\frac{1}{p_k} - \frac{1}{p_{k-1}} \right) & \qquad k > 0\ \ [\text{A sector}], \\
                 &\frac{p_0}{p_0} = 1                                                 & \qquad k = 0\ \ [\text{L sector}].
               \end{aligned}
               \right.
\end{equation}
Note that in the study of spin glasses the matrix $A^{ab;cd}$ is the Hessian of the fluctuations, and
the diagonal elements $a\cap b = R+1$ and $c\cap d = R+1$ are identically zero. 
The values $r=R+1$ and $k=R+1$ must then be excluded reducing the number of 
block  to $R+1$ and their size to $(R+1)\times (R+1)$.

The  eigenvector $\phi^{r}_k(\hat{a},\hat{b}) =  \phi^r_{\hat{k}} \psi^{r}_k(\hat{a},\hat{b};\hat{e})$ 
corresponds  to the eigenvector $\phi^{r}_k(a,b)=  \phi^r_{\hat{k}} \psi^{r}_k(a,b;e)$  
in the Replica Space with components [cfr. eq. (\ref{eq:eigenvect})]:
\begin{equation}
 \phi^r_t = \left\{
            \begin{aligned}
              0                                                                                           & \qquad t < k - 1,\\
              -\frac{1}{p^{(r)}_{k-1}}\, \phi^r_{\hat{k}}                                           & \qquad t = k -1, \\
              \left(\frac{1}{p^{(r)}_k} - \frac{1}{p^{(r)}_{k-1}}\right)\, \phi^r_{\hat{k}} & \qquad t > k -1.
            \end{aligned}
          \right.
\end{equation}
that are {\sl null} when the {\sl cross}-overlap $t$ is smaller that the resolution $k$ of the RFT, and
independent of $t$ when larger than $k$.
The functions  
\begin{equation}
                   \psi^r_k(a,b;e) = 
                         \frac{1}{2}\bigl[  \pbar_k(a,e) +  \pbar_k(b,e)
                         \bigr]\, \Delta_r(a,b),        
\end{equation}
form the  first and second family of basis vectors of Ref. \cite{Temal94}. 
In our notation 
$| k; r; c_0,\dotsc, c_{k-1}\rangle \propto \psi^r_k(a,b;c)$ where $c_0,\dotsc,c_{k-2}$ are 
the label ``$a$" and $c_{k-1}$ the label ``$b$" of Ref \cite{Temal94} .
For example the vector $|1;0,b\rangle$ of Ref \cite{Temal94} is
\begin{equation}
  | 1; 0, c_0\rangle = C\, \psi^0_1(a,b;c) =  \left\{ 
   \begin{aligned} 
        A  &= -\frac{C}{p_0} 
                        & (a_0\not=c_0, b_0\not=c_0)\\
	B &= C \left(\frac{1}{2p_1} - \frac{1}{p_0}\right)
	 & \begin{aligned} &(a_0\not=c_0, b_0 = c_0) \\  &(a_0=c_0, b_0\not=c_0)\end{aligned}
   \end{aligned}
   \right.
\end{equation}
where $C=-p_0$  to have $A=1$, and that $B=(1/2)(2 - p_0/p_1)$.

%Note also that due to the different notation, the first family $\psi_k(a,b)$ 
%is recovered for $r=R+1$,  in agreement with the equality 
%$A^{R+1}(a,b;c)= A^{R+1}(a,a;c) =  A^{ac}$.

\bigskip
\noindent
$\bullet$ {\sl The Replicon sector} \\

The functions $\psi^{\, r,r}_{k,l}(\hat{a}, \hat{b}; \hat{c}, \hat{d})$ do not obey a sum rule 
similar to, e.g.,  Eq. (\ref{eq:s-psikl-s-psikl}),  and hence are not
eigenvectors. 
From this sum rule, and Eq. (\ref{eq:4rep-quad}), it follows, however, that 
$\phi^{r}_{k,l}(\hat{a},\hat{b})= 
\overline{\phi}^{\, r,r}_{\hat{k},\hat{l}}\, \overline{\psi}^{\, r,r}_{k,l}(\hat{a},\hat{b};\hat{e},\hat{f})$
diagonalizes the eigenvalue equation in the Replicon sector.
This choice leads to eigenvectors in the Replicon sector symmetrical in both the first and 
second pair  of indexes. 
In Ref. \cite{Temal94} the symmetry in the second pair of indexes was not imposed, 
for eigenvectors only symmetry in the first two indexes is mandatory. Then to make contact with 
the eigenvectors of Ref. \cite{Temal94},
we introduce the new set functions, 
\begin{equation}
\label{eq:tilde-psikl}
\begin{split}
  &\widetilde{\psi}^{\, r,r}_{k,l}(\hat{a}, \hat{b}; \hat{c}, \hat{d}) = 
     \varphi^{r,r}_{k,l}(\hat{a}, \hat{b}; \hat{c}, \hat{d}) \\
     &\phantom{=====}
   - \sum_{t=0}^{R+1}\frac{4}{\delta_r^{(t-1)}}\, 
         \Deltastar_{k;r+1}(\hat{c},0)\, \Deltastar_{l;r+1}(\hat{d},0)\,
     \overline{\psi}^{\,r,r}_t(\hat{a},\hat{b}; \hat{c},\hat{d}),
\end{split}     
\end{equation}
satisfying the identity:
\begin{equation}
\label{eq:t-psikl-t-psik}
  \sum_{\hat{a},\hat{b}}  \widetilde{\psi}^{\, r,r}_{u,v}(\hat{a},\hat{b};\hat{c}, \hat{d})\,  
                                         \psi^{s}_k(-\hat{a},-\hat{e};\hat{e})
                                         = 0,
\end{equation}
see Eq. (\ref{eq:Qv11}).
These functions obey the sum rule
\begin{equation}
\label{eq:psikl-t-psikl}
  \sum_{\hat{e},\hat{f}}  \psi^{\, r,r}_{u,v}(\hat{a},\hat{b};\hat{e}, \hat{f})\,  
                                         \widetilde{\psi}^{\, s,s}_{k,l}(-\hat{e},-\hat{f};\hat{c}, \hat{d})
                                         = \delta_{r,s}\, \delta_{(u,v)(k,l)}\,
                                         \widetilde{\psi}^{\, r,r}_{k,l}(\hat{a},\hat{b};\hat{c}, \hat{d}).                
\end{equation}
and hence 
$\phi^{r}_{k,l}(\hat{a},\hat{b})= 
\phi^{\, r,r}_{\hat{k},\hat{l}}\, \widetilde{\psi}^{\, r,r}_{k,l}(\hat{a},\hat{b};\hat{e},\hat{f})$
are eigenvectors in the Replicon sector.
The functions  $\widetilde{\psi}^{\, r,r}_{k,l}(\hat{a},\hat{b};\hat{e},\hat{f})$ 
correspond to the third family of eigenvectors  introduced in  Ref. \cite{Temal94}.
It is straightforward to show  
that the two sets of eigenvectors are related by
$\overline{\psi}^{\, r,r}_{k,l}(\hat{a},\hat{b};\hat{e},\hat{f}) = 
[\widetilde{\psi}^{\, r,r}_{k,l}(\hat{a},\hat{b};\hat{e},\hat{f}) + 
 \widetilde{\psi}^{\,r,r}_{k.l}(\hat{a},\hat{b};\hat{f},\hat{e})]/2$.

In the Replicon sector the RFT diagonalizes completely  the eigenvalue equation 
with eigenvalues
\begin{equation}
\label{eq:eigenR}
  \Lambda = A^{r,r}_{\hat{k},\hat{l}},
  \qquad k,l \geq r+1,
\end{equation}
and each eigenvalue has multiplicity:
\begin{equation}
  \mu(k,l) = \frac{p_0}{2} \bigl[ p_{r} - (1+\delta_{k,r+1} + \delta_{l,r+1})\, p_{r+1} \bigr]\,
                       \overline{\delta}_k\, \overline{\delta}_l
\end{equation}
where
\begin{equation}
  \overline{\delta}_k  = 
               \left\{
               \begin{aligned}
                 & \frac{1}{p_k} - \frac{1}{p_{k-1}}  & \qquad k > r+1, \\
                 &\frac{1}{p_{r+1}}                                & \qquad k = r+1.
               \end{aligned}
               \right.
\end{equation}

In the Replica Space the eigenvalue equation in the Replicon sector takes the form:
\begin{equation}
    \frac{1}{2}\sum_{cd} A^{ab;cd}\, \phi^{cd;ef} = \Lambda\, \phi^{ab;ef},
\end{equation}
where $e$ and $f$ are two passive replicas with $e\cap f = a\cap b = r = 0,\dotsc, R$, and
eigenvectors
$\phi^{r}_{k,l}(a,b)= \phi^{r,r}_{\hat{k},\hat{l}}\, \widetilde{\psi}^{\, r,r}_{k,l}(a,b;e,f)$ with 
$\widetilde{\psi}^{\, r,r}_{k,l}(a,b;e,f)$  the inverse RFT of 
 $\widetilde{\psi}^{\, r,r}_{k,l}(\hat{a},\hat{b};\hat{e},\hat{f})$. 
For $k,l> r+1$ the read:
\begin{equation}
  \widetilde{\psi}^{\, r,r}_{k,l}(a,b;e,f) = \bigl[ \pbar_k(a,e)\,\pbar_l(b,f) 
            + \pbar_k(b,e)\, \pbar_l(a,f)\bigr]\,
                  \Delta_r(a,b),
\end{equation}
while for the boundary case  $k>r+1$ and $l=r+1$ [or $k=r+1$ and $l>r+1$]: 
\begin{equation}
\begin{split}
  \psi^{r,r}_{k,r+1}(a,b;e,f) &=  \Bigl\llbracket
                               \frac{1}{p_{r+1}} 
                               \bigl[ \pbar_k(a,e)\,\delta_{r+1}(b,f) + \pbar_k(b,e)\, \delta_{r+1}(a,f)\bigr]\\
                    &\phantom{=}
                    - \frac{1}{p_r - p_{r+1}} 
                               \bigl[ \pbar_k(a,e) + \pbar_k(b,e)\bigr]\,\Delta_r(e,f)
              \Bigr\rrbracket\,
                  \Delta_r(a,b).
\end{split}                  
\end{equation}
and finally for $k=l=r+1$:
\if 0
\begin{equation}
\begin{split}
  \psi^{r,r}_{r+1,r+1}&(a,b;e,f) =  
                               \frac{1}{p_{r+1}^2} 
                               \bigl[ \delta_{r+1}(a,e)\,\delta_{r+1}(b,f) + \delta_{r+1}(b,e)\, \delta_{r+1}(a,f)\bigr]
                               \,\Delta_r(a,b)\\
                    &\phantom{}
                    - \Bigl\llbracket \frac{1}{p_r -2 p_{r+1}} 
                               \bigl[ \pbar_{r+1}(a,e)+\pbar_{r+1}(b,f) + \pbar_{r+1}(b,e)+ \pbar_{r+1}(a,f)\bigr]\\
                    &\phantom{}                               
                    -  \frac{1}{2p_{r}(p_r - p_{r+1})}
                                                   \bigl[ \delta_{r}(a,e)+\delta_{r}(b,f) + \delta_{r}(b,e)+ \delta_{r}(a,f)\bigr]
              \Bigr\rrbracket
                     \\
                     &\phantom{=========================}
                     \times                              
              \Delta_r(e,f)\,
                  \Delta_r(a,b).
\end{split}                  
\end{equation}
\fi
\begin{equation}
\begin{split}
  &\psi^{r,r}_{r+1,r+1}(a,b;e,f) =  
            \left\llbracket 
                               \frac{1}{p_{r+1}^2} 
                               \bigl[ \delta_{r+1}(a,e)\,\delta_{r+1}(b,f) + \delta_{r+1}(b,e)\, \delta_{r+1}(a,f)\bigr]
                               %\,\Delta_r(a,b)
                              \right. \\
                    &\phantom{--}
                    - \frac{1}{p_{r+1}(p_r -2 p_{r+1})} 
                               \bigl[ \delta_{r+1}(a,e)+\delta_{r+1}(b,f) + \delta_{r+1}(b,e)+ \delta_{r+1}(a,f)\bigr]\\
                    &\phantom{---}                               
                    \left.
                    +  \frac{1}{2(p_{r}-p_{r+1})(p_r - 2p_{r+1})}
                                                   \bigl[ \delta_{r}(a,e)+\delta_{r}(b,f) + \delta_{r}(b,e)+ \delta_{r}(a,f)\bigr]
              \right\rrbracket
                     \\
                     &\phantom{=====================}
                     \times                              
              \Delta_r(a,b)\,
                  \Delta_r(e,f).
\end{split}                  
\end{equation}
The reader is referred to Ref. \cite{Temal94} for a detailed discussion of the structure
of  $\tilde{\psi}^{r,r}_{k,l}(a,b;e,f)$.

The total dimension of the LA sector is $p_0(R+2)$, see Eq. (\ref{eq:dimLAr}),  while that of 
the Replicon Sector is:
\begin{equation}
 \sum_{r=0}^{R} \mbox{Dim}_r[\psi^{r,r}_{k,l}]  = \frac{1}{2}p_0\, (p_0 - 2 R - 3)
\end{equation}
Taken all together one recovers the total dimension of $p_0(p_0+1)/2$ of the 
space of  the symmetric two-replicas functions $\phi^{ab}$.\footnote{
If the diagonal terms are not included the dimension is reduces by a $p_0$ 
leading to $p_0(p_0-1)/2$.}

\section{Gaussian Fluctuations in the Spherical Model}
\label{sec:SphMod}

To motivate the calculation of the eigenvalues of the functions
$A^{ab;cd}$ discussed in the previous section, here we show how a matrix of this form 
arises naturally
in the study of the static properties of spin glass model.
We shall consider the case of the generic spin glass spherical model
where  $A^{r,s}_t$ and $A^{r,r}_{u,v}$ can be computed explicitly.
The model  is described by the Hamiltonian,
\begin{equation}
  {\cal H} = -\sum_{p\geq 2}\ \sum_{i_1 < \dotsb < i_p}^{1,N}\, J^{(p)}_{i_1,\dotsc,i_p}
                                  \sigma_{i_1}\dotsm\sigma_{i_p},
\end{equation}
where $\sigma_i$ are $N$ continuous real variables which range from $-\infty$ to $+\infty$
subject to the global spherical constraint $\sum_{i=1}^{N}\sigma_i^2 = N$. The couplings
$J^{(p)}_{i_1,\dotsc,i_p}$ are quenched, independent identical distributed Gaussian variables of
zero mean and variance $\overline{(J^{(p)}_{i_1,\dotsc,i_p})^2}= p! J_p^2 / 2 N^{p-1}$.
The scaling with $N$ ensures an extensive free energy so that the thermodynamic limit $N\to\infty$
is well defined. We do not consider the presence of an external magnetic field, the extension is
however straightforward.

The properties of the model depend on the number and the value of the $p$'s in the sum
\cite{CriLeu04,CriLeu06,CriLeu07,Sunetal12,CriLeu13}.
We shall not discuss this issue here, since we are only interested into the general expression 
of  the eigenvalues controlling the stability of saddle point controlling the static, thermodynamic,
properties of the model in the thermodynamic limit $N\to\infty$.

The static properties of the model follow from the minimum of the quenched 
free-energy functional obtained from the logarithm of the partition function averaged 
over the couplings distribution.
A direct calculation of the average is a difficult task, and this is evaluated using the {\sl replica
trick}. That is, one introduces $n$ identical replicas of the system and compute the 
{\sl annealed} average, i.e., the average of the partition function of the replicated system. 
The sought quenched 
free energy functional is obtained from the $n\to 0$ limit of the annealed free-energy functional.
The interested reader can find details of the calculation for the spherical model in 
Ref. \cite{CriSom92,CriLeu06}.

When this program is carried on, and the thermodynamic limit $N\to\infty$ is taken, 
one ends up with a free energy density $f$ given by 
\begin{equation}
\label{eq:free-en}
  \beta f = \lim_{n\to 0}\frac{1}{n}\,  G[Q] \Big\rvert_{\mbox{sp}} + \mbox{constant}
\end{equation}
where $\beta = 1/T$ is the inverse temperature and  
\begin{equation}
  G[Q] = -\frac{1}{2} \sum_{a,b}^{1,n} g(Q^{ab}) -\frac{1}{2}\mbox{Tr} \ln Q,
\end{equation}
with
\begin{equation}
  g(x)  = \sum_{p\geq 2} \frac{\mu_p}{p}x^p, 
  \qquad \mu_p = (\beta J_p)^2p/2,
\end{equation}
is a functional of the symmetric overlap matrix 
$Q^{ab} = \overline{\langle\sigma_i^a\sigma_i^b\rangle}$
between replicas $a$ and $b$, brackets denote thermal average. 
Equation (\ref{eq:free-en}) follows from a saddle point calculation valid for $N\to\infty$, and
the functional $G[Q]$ in Eq. (\ref{eq:free-en}) is  evaluated at its stationary point (saddle point):
\begin{equation}
\label{eq:stpt}
 \frac{\delta}{\delta Q^{ab}} G[Q] = - \Lambda\bigl(Q^{ab}\bigr) - (Q^{-1})^{ab} = 0,
 \qquad a < b,
\end{equation}
where $\Lambda(x) = d g(x)/ dx = g'(x) = \sum_{p\geq 2} \mu_p x^{p-1}$.
The diagonal elements $Q^{aa}$ are fixed by the spherical constraint and read
$Q^{aa} = 1$.

Stability of the saddle point  requires that the all eigenvalues of the Gaussian
fluctuations for about the stationary point be non negative for $n\to 0$. 
The dimension of $Q^{ab}$ is $(1/2)n(n-1)$ and is negative for $n\to 0$, changing minima into maxima.
The Gaussian fluctuations are described by Hessian
\begin{equation}
\begin{split}
 A^{ab;cd} &=  \frac{\delta}{\delta Q^{ab}}   \frac{\delta}{\delta Q^{cd}} G[Q] \\
                   &= -\Lambda'(Q^{ab})\, \delta_{(ab),(cd)} 
                      + (Q^{-1})^{ac} (Q^{-1})^{bd}
                      + (Q^{-1})^{ad} (Q^{-1})^{bc},
\end{split}                      
\end{equation}
with $a<b$ and $c<d$, evaluated at the stationary point Eq. (\ref{eq:stpt}).

In a scenario with R replica symmetry breaking steps the matrix
$Q^{ab}$ has the form (\ref{eq:matrix}) with $A_r = Q_r$ and, similarly
\begin{equation}
 (Q^{-1})^{ab} = \sum_{r=0}^{R+1} Q^{-1}_r\, \Delta_r(a,b)
\end{equation}
with $(Q^{-1})^{ab}\rvert_{a\cap b = r} = Q^{-1}_r$ given by:
\begin{equation}
  Q^{-1}_r = \sum_{k=0}^{r} \frac{1}{p_k} 
         \left[\frac{1}{Q_{\hat{k}}} - \frac{1}{Q_{\widehat{k+1}}} \right],
\end{equation}
where $Q_{\hat{k}}$ is the RFT of $Q_r$,  see Eq. (\ref{eq:invmatrix}).

The LA $A^{r,s}_t$ and Replicon $\leftidx{_{\rm R}}A^{r,r}_{u,v}$ contributions can be 
evaluated directly from their 
tree representation. Because of the constraint $a < b$ and $c < d$, only topologically 
different trees, i.e., the ones that cannot be obtained one from the other by exchanging 
$a$ and $b$ or $c$ and $d$, must be considered.
Assume $r>s$, then the first two diagrams  of the first row and the first diagram of the second 
row of Fig. \ref{fig:4replica-rgs} gives for the LA contribution:
\begin{equation}
  A^{r,s}_t = \left\{\begin{aligned}
                                  & 2 Q^{-1}_t\,Q^{-1}_t & \qquad t\leq s < r,\\
                                   &2 Q^{-1}_t\,Q^{-1}_s & \qquad s< t \leq  r,\\
                                   &Q^{-1}_t\,Q^{-1}_s + Q^{-1}_r\,Q^{-1}_s & \qquad s < r < t.                           
                              \end{aligned}
                     \right.
\end{equation} 
When $r=s$ from the first two diagrams of Fig. \ref{fig:4replica-res} one has:
\begin{equation}
  A^{r,r}_t = \left\{\begin{aligned}
                                  & 2 Q^{-1}_t\,Q^{-1}_t & \qquad t\leq s = r,\\
                                   &Q^{-1}_t\,Q^{-1}_r + Q^{-1}_r\,Q^{-1}_r & \qquad s = r < t.                           
                              \end{aligned}
                     \right.
\end{equation}
if $r < R+1$, and  
\begin{equation}
  A^{R+1,R+1}_{R+1}  = -\Lambda'(Q_{R+1}) + 2 Q^{-1}_{R+1}\,Q^{-1}_{R+1}, 
\end{equation}
where $Q_{R+1} = Q^{aa} = 1$, if $r=s=t=R+1$.

The Replicon contribution is obtained from the diagram of Fig. \ref{fig:4replica-rep} and reads:
\begin{equation}
\label{eq:R1RSB}
  \leftidx{_{\rm R}}A^{r,r}_{u,v} = -\Lambda'(Q_r)\,\delta_{u,R+1}\,\delta_{v,R+1} 
                           + Q^{-1}_u\,Q^{-1}_v + Q^{-1}_r\,Q^{-1}_r.
\end{equation}

It is now straightforward to evaluate the RFT. 
Assume as before that $ r > s$ then,  with the above form of $A^{r,s}_t$, from 
Eq. (\ref{eq:LARFT}) one finds for $s < r < k$ :
\begin{equation}
\begin{split}
 A^{r,s}_{\hat{k}} &= 4 Q^{-1}_s \sum_{t=k}^{R+1} p_t\bigl[ Q^{-1}_t - Q^{-1}_{t-1}\bigr]\\
                              &= 4 Q^{-1}_s Q^{-1}_{\hat{k}} 
                                 = 4 \frac{Q^{-1}_s}{Q_{\hat{k}}}.
\end{split}                              
\end{equation}
The last equality follows from  $Q^{-1}_{\hat{k}} = 1/Q_{\hat{k}}$, see Eq. (\ref{eq:invmatRFT}).
This results does not depend on $r$, provided it is larger than $s$,  
and it is indeed valid also for $s < k \leq r$, as the direct calculation shows.
For $k \leq s < r$ the RFT can be written as
\begin{equation}
 A^{r,s}_{\hat{k}} = \sum_{t=k}^{s} p^{(r,s)}_t \bigl[ A^{r,s}_t - A^{r,s}_{t-1}\bigr]
                                  + A^{r,s}_{\widehat{s+1}},
\end{equation}
so that in this range of $k$
\begin{equation}
 A^{r,s}_{\hat{k}} = 2 \sum_{t=k}^{s} p_t \bigl[ (Q^{-1}_t)^2 - (Q^{-1}_{t+1})^2\bigr]
                                  + 4 \frac{Q^{-1}_s}{Q_{\widehat{s+1}}}.
\end{equation}
The case $r=s$ can be worked out in a similar way. 
The results can be combined and the RFT of $A^{r,s}_t$ for generic $r$, $s$ and $k$ can be
 written as
 \begin{equation}
 \label{eq:LARFTsp}
  A^{r,s}_{\hat{k}} = \left\{\begin{aligned}
                                  & 4\, \frac{Q^{-1}_m}{Q_{\hat{k}}} & \qquad k > m,\\
                                   &2 \sum_{t=k}^{m} p_t \bigl[ (Q^{-1}_t)^2 - (Q^{-1}_{t+1})^2\bigr]
                                     +  4\, \frac{Q^{-1}_m}{Q_{\widehat{m+1}}} & \qquad k \leq m,                           
                              \end{aligned}
                     \right.
\end{equation} 
where $m = \min(r,s)$.

The RFT of $A^{r,r}_{u,v}$ is given by  Eq. (\ref{eq:dblRFT}) which, with Eq. (\ref{eq:R1RSB}),
gives:
\begin{equation}
 \label{eq:RRFTsp}
\begin{split}
  A^{r,r}_{\hat{k},\hat{l}} &= -\Lambda'(Q_r) 
                                            + \sum_{u=k}^{R+1} \sum_{u=l}^{R+1}
                                            p_u p_v 
                                            \bigl[ Q^{-1}_u - Q^{-1}_{u+1}\bigr]\,
                                            \bigl[ Q^{-1}_v - Q^{-1}_{v+1}\bigr] \\
                                           &= -\Lambda'(Q_r) + \frac{1}{Q_{\hat{k}}\, Q_{\hat{l}}},
                                           \qquad\qquad k,l \geq r+1.
\end{split}                                            
\end{equation}

The eigenvalues of the Gaussian fluctuations about the saddle point are obtained from 
Eqs. (\ref{eq:eigenLA}) and (\ref{eq:eigenR}) using the above expression for 
$A^{r,s}_{\hat{k}}$ and $A^{r,r}_{\hat{k},\hat{l}}$ evaluated at the stationary point 
$\Lambda(Q_r) = Q^{-1}_r$ of $G[Q]$, see Eq. (\ref{eq:stpt}).
In the case $R=1$, i.e.,  only one step of replica symmetry breaking,  one  recovers 
the results of  Ref. \cite{CriSom92}. Some details can be found in \ref{app:Fluct1RSB}.

\section{Conclusion}
In this work we have done a detailed discussion of the RFT.  The use of RFT has been illustrated 
by computing explicitly the RFT of some typical function of the Replica Space.
We have also shown how to use RFT to find the eigenvalues of matrices of the form $A^{ab;cd}$
which appear in the study of Gaussian fluctuations about the saddle point in spin glass models.
The results are not new and we reproduces the results of Ref. \cite{Temal94}.
In this Reference the diagonalization was performed by a direct construction of the eigenvectors  
in the Replica Space. The procedure is rather involved.  Here we present the solution using the
RFT which, to our knowledge, was never presented in detail.
Finally we have computed the eigenvalue of the Gaussian fluctuations for a generic spin glass 
spherical model generalizing the results of Ref. \cite{CriSom92} to an arbitrary number of 
replica symmetry breaking.

\bigskip
\noindent {\em Acknowledgments ---}
We are indebted  to T. Temesv\'ari for critical reading of the manuscript and helpful suggestions.
AC acknowledge financial support from the European Research
Council through ERC grant agreement no. 247328

\appendix
\section{Derivation of Eq. (\ref{eq:4repLARFT}).}
\label{app:4repLARFT}
The starting point is  Eq. (\ref{eq:4rep-LA}). This is composed by the sum of four terms 
which can be obtained 
one from the other by exchanging $a$ and $b$ or $c$ and $d$. So we can consider only the first
 term proportional to
 \begin{equation}
   \delta_t(a,c)\,\Delta_r(a,b)\,\Delta_s(c,d),
 \end{equation}
 the others three being obtained with suitable permutations of indexes.
 The RFT is evaluated  in sequence on each index. 
By using Eq. (\ref{eq:pstar_1}) the RFT on index $d$ leads to:
\begin{equation}
 \sum_d (\dotso)\,\chi(d\hat{d}) =   \delta_t(a,c)\,\Delta_r(a,b)\, \pstar_s(\hat{d},0)\,\chi(c\hat{d}).
\end{equation}
By using now Eqs. (\ref{eq:prod}) and (\ref{eq:deltachi}) the RFT on index $c$ gives:
\begin{equation}
 \sum_c (\dotso)\,\chi(c\hat{c}) =   \frac{1}{\sqrt{p_0}}\, 
                                                           p_t \deltastar_t(\hat{c}+\hat{d}, 0)\,
                                                           \Delta_r(a,b)\, 
                                                           \pstar_s(\hat{d},0)\,\chi\bigl(a(\hat{c}+ \hat{d})\bigr).
\end{equation}
The RFT on index $b$ is evaluated using Eqs. (\ref{eq:prod}) and (\ref{eq:pstar_1}):
\begin{equation}
 \sum_b (\dotso)\,\chi(b\hat{b}) =   \frac{p_t}{p_0}\, 
                                                           \deltastar_t(\hat{c}+\hat{d}, 0)\,
                                                           \pstar_r(\hat{b},0)\, 
                                                           \pstar_s(\hat{d},0)\,\chi\bigl(a(\hat{b}+\hat{c}+ \hat{d})\bigr).
\end{equation}
Finally the RFT on index $a$ is evaluated using Eq. (\ref{eq:inv}) and produces the delta function 
ensuring the vanishing of the total momentum:
\begin{equation}
 \sum_a (\dotso)\,\chi(a\hat{a}) =   \frac{p_t}{p_0}\, 
                                                           \deltastar_t(\hat{c}+\hat{d}, 0)\,
                                                           \pstar_r(\hat{b},0)\, 
                                                           \pstar_s(\hat{d},0)\,
                                                           \delta_{R+1}(\hat{a} + \hat{b}+\hat{c}+ \hat{d}, 0).
\end{equation}
The other three terms are obtained by exchanging $a\leftrightarrow b$  and  $c\leftrightarrow d$.
Taken all together these leads to:
\begin{equation}
\begin{split}
  \leftidx{_{\rm LA}}A^{r,s}(\hat{a},\hat{b};\hat{c},\hat{d}) = \frac{1}{4p_0}
                                    \sum_{t=0}^{R+1}  p_t^{(r,s)} 
                                      &\bigl[ A^{r,s}_t - A^{r,s}_{t-1}\bigr] \, \deltastar_t(\hat{c}+\hat{d},0)\\
                                               &   
                                   \times
                                          \bigl[\pstar_r(\hat{a},0) + \pstar_r(\hat{b},0)\bigr]
                                          \bigl[\pstar_s(\hat{c},0) + \pstar_s(\hat{d},0)\bigr]\\
    &   
           \times
                 \delta_{R+1}(\hat{a}+\hat{b}+\hat{c}+\hat{d},0).
\end{split}
\end{equation}
Introducing now $A^{r,s}_{\hat{t}}$ from Eq. (\ref{eq:4repRFTLA}), and using 
the identity
\begin{equation}
  \sum_{t=0}^{R+1} \bigl[f_{\hat{t}} - f_{\widehat{t+1}}\bigr]\, \deltastar_t(\dotso)
   =  \sum_{t=0}^{R+1} f_{\hat{t}}\, \Deltastar_t(\dotso),
\end{equation}
Eq. (\ref{eq:4repLARFT}) follows (QED).

\section{Derivation of Eq. (\ref{eq:4repRRFT}).}
\label{app:4repRRFT}
The starting point is  Eq. (\ref{eq:4rep-R}), and the RFT is evaluated following the same
procedure of \ref{app:4repLARFT}.
Consider the first term proportional to
 \begin{equation}
   \delta_u(a,c)\,\delta_v(b,d)\,\Delta_r(a,b)
 \end{equation}
 the other follows  by exchanging  $c$ with $d$. 
By using Eq. (\ref{eq:deltachi}) the RFT on index $d$ leads to:
\begin{equation}
 \sum_d (\dotso)\,\chi(d\hat{d}) =  \delta_u(a,c)\, 
                                                           p_v\,\deltastar_v(\hat{d},0)\,  \chi(b\hat{d})\,
                                                           \Delta_r(a,b),
\end{equation}
and that on index $c$ to
\begin{equation}
 \sum_c (\dotso)\,\chi(c\hat{c}) =   p_u\,\deltastar_u(\hat{c},0)\,  \chi(a\hat{c})\,
                                                           p_v\,\deltastar_v(\hat{d},0)\,  \chi(b\hat{d})\,
                                                           \Delta_r(a,b),
\end{equation}
The RFT on index $b$ is evaluated using Eqs. (\ref{eq:prod}) and (\ref{eq:pstar_1}):
\begin{equation}
 \sum_b (\dotso)\,\chi(b\hat{b}) =   \frac{p_u p_v}{p_0}\,
                                                            \deltastar_u(\hat{c},0)\,  
                                                           \deltastar_v(\hat{d},0)\,  
                                                           \pstar_r(\hat{b}+\hat{d},0 )\,
                                                           \chi\bigl(a(\hat{b}+\hat{c}+ \hat{d})\bigr).
 \end{equation}
Finally the RFT on index $a$  leads to the delta function 
ensuring the vanishing of the total momentum:
\begin{equation}
 \sum_a (\dotso)\,\chi(a\hat{a}) =   \frac{p_u p_v}{p_0}\,
                                                            \deltastar_u(\hat{c},0)\,  
                                                           \deltastar_v(\hat{d},0)\,  
                                                           \pstar_r(\hat{b}+\hat{d},0 )\,
                                                           \delta_{R+1}(\hat{a} + \hat{b}+\hat{c}+ \hat{d}, 0).
 \end{equation}
By adding the second term the RFT of the Replicon contribution reads:
\begin{equation}
\begin{split}
 \leftidx{_{\rm R}}A^{r,r}(\hat{a},\hat{b};\hat{c},\hat{d}) &= 
       \frac{1}{p_0}\,
           \sum_{u=r+1}^{R+1}\sum_{v=r+1}^{R+1} 
                     p_u p_v  \llbracket A\rrbracket^{r,r}_{u,v}\,
                           \deltastar_u(\hat{c},0)\, \deltastar_v(\hat{d},0)
                           \\
    & \phantom{=}
     \times
       \bigl[\pstar_r(\hat{a}+\hat{c},0) + \pstar_r(\hat{a}+\hat{d},0)\bigr] 
       \delta_{R+1}(\hat{a}+\hat{b}+\hat{c}+\hat{d},0).
\end{split}
\end{equation}
By introducing the RFT $A^{r,r}_{\hat{u},\hat{v}}$ of $A^{r,r}_{u,v}$ from relation 
(\ref{eq:dblRFT1}), and using two times the identity
\begin{equation}
  \sum_{u=r+1}^{R+1} \bigl[f_{\hat{u}} - f_{\widehat{u+1}}\bigr]\, \deltastar_u(\dotso)
   =  \sum_{u=r+1}^{R+1} f_{\hat{u}}\, \Deltastar_{u,r+1}(\dotso),
\end{equation}
where $\Deltastar_{u,r+1}(\dotso)$ is defined in Eq. (\ref{eq:deltastar_1}),
to disentangle the RFT components $A^{r,r}_{\hat{u},\hat{v}}$,
Eq. (\ref{eq:4repRRFT}) follows (QED).

\section{Derivation of Eqs. (\ref{eq:Qv1}) and (\ref{eq:Qv11}).}
\label{app:Qv1}
The sum over $\hat{d}$  in Eq. (\ref{eq:Qv1}) is done by exploiting the 
$\delta_{R+1}$ functions, see Eqs. (\ref{eq:psik}) and (\ref{eq:varphikl}), and leads to
\begin{equation}
\begin{split}
  \sum_{\hat{c},\hat{d}}  &\varphi^{r,r}_{u,v}(\hat{a},\hat{b};\hat{c}, \hat{d})\,  
                                         \psi^{s}_k(-\hat{c},-\hat{d};\hat{e}) \\
                                       &= \frac{1}{2 p_0^{3/2}}
                                       \sum_{\hat{c}}
                                       \bigl[\pstar_r(\hat{a}+\hat{c},0) + \pstar_r(\hat{b}+\hat{c},0)\bigr]
                                       \bigl[\pstar_s(\hat{c},0) + \pstar_s(\hat{c}-\hat{e},0)\bigr] \\
                                       &\phantom{===}\times
                                       \Deltastar_{u;r+1}(\hat{c},0)\,
                                       \Deltastar_{v;r+1}(\hat{c}-\hat{e},0)\,
                                       \Deltastar_{k}(\hat{e},0)\,
				\delta_{R+1}(\hat{a}+\hat{b}+\hat{e},0).
\end{split}                                             
\end{equation}
The products are evaluated using the following identities valid for $u,v \geq r+1$:
%\begin{equation}
\begin{align}
  \Deltastar_{u;r+1}(\hat{c},0)\,\pstar_r(\hat{c},0) &= \delta_{u,r+1}\,\pstar_r(\hat{c},0),
\\
  \pstar_r(\hat{c},0)\,\Deltastar_{v;r+1}(\hat{c},\hat{e})\,\Deltastar_k(\hat{e},0) &= 
  \delta_{v,\max(r+1, k)}\,\pstar_r(\hat{c},0)\, \Deltastar_k(\hat{e},0),
%  \quad m = \max(r+1, k),
\end{align}  
%\end{equation}
so that
\begin{equation}
  \Deltastar_{u;r+1}(\hat{c},0)\, \Deltastar_{v;r+1}(\hat{c},\hat{e})\,  
    \pstar_r(\hat{c},0)\ \Deltastar_k(\hat{e},0) =
       \delta_{u,r+1}\,  \delta_{v,m}\, \pstar_r(\hat{c},0)\, \Deltastar_k(\hat{e},0),
\end{equation}
and
\begin{equation}
  \Deltastar_{u;r+1}(\hat{c},0)\, \Deltastar_{v;r+1}(\hat{c},\hat{e})\,  
    \pstar_r(\hat{c},\hat{e})\ \Deltastar_k(\hat{e},0) =
       \delta_{u,m}\,  \delta_{v,r+1}\, \pstar_r(\hat{c},\hat{e})\, \Deltastar_k(\hat{e},0),
\end{equation}
where $m= \max(r+1, k)$.
The sum over $\hat{c}$ can now be performed by using Eq. (\ref{eq:pstar_pstar}).
For example
\begin{equation}
\begin{split}
    \sum_{\hat{c}}
                                      \bigl[\pstar_r(\hat{a}&+\hat{c},0) + \pstar_r(\hat{b}+\hat{c},0)\bigr]
                                       \pstar_s(\hat{c},0)
                                       \Deltastar_{u;r+1}(\hat{c},0)\,
                                       \Deltastar_{v;r+1}(\hat{c}-\hat{e},0)\,
                                       \Deltastar_{k}(\hat{e},0)
                                       \\
                                &=       \delta_{u,r+1}\, \delta_{v,m}\, 
                                       \Deltastar_{k}(\hat{e},0)
                                    \sum_{\hat{c}}
                                      \bigl[\pstar_r(\hat{a}+\hat{c},0) + \pstar_r(\hat{b}+\hat{c},0)\bigr]
                                      \pstar_s(\hat{c},0)  \\                                     
                                &=     p_0\, \delta_{r,s}\,  \delta_{u,r+1}\, \delta_{v,m}\, 
                                       \Deltastar_{k}(\hat{e},0)\,
                                      \bigl[\pstar_r(\hat{a},0) + \pstar_r(\hat{b},0)\bigr].
\end{split}                                       
\end{equation}
%Then,  with the contribution from $\pstar_s(\hat{c}-\hat{e},0)$,
By adding the contribution from $\pstar_s(\hat{c}-\hat{e},0)$, and rearranging the terms,
Eq. (\ref{eq:Qv1}) follows.  (Q.E.D)

Equation (\ref{eq:Qv11}) is obtained following a similar procedure and evaluating the 
$\pstar_r(\dotsc) \pstar_s(\dotsc)$ products with the help of Eq. (\ref{eq:pstar_pstar}).

\section{Derivation of Eq. (\ref{eq:s-psikl-s-psikl}).}
\label{app:psikl-psikl}

Consider the case $r=s$, then:
\begin{equation}
\begin{split}
  \sum_{\hat{c},\hat{d}}  \overline{\psi}^{\,r,r}_{u,v}(\hat{a},\hat{b};\hat{c}, \hat{d})\,  
                                        & \overline{\psi}^{\, r,r}_{k,l}(-\hat{c},-\hat{d};\hat{e}, \hat{f})
                                         =   \sum_{\hat{c},\hat{d}} 
                                          \overline{\psi}^{\,r,r}_{u,v}(\hat{a},\hat{b};\hat{c}, \hat{d})\,  
                                         \overline{\varphi}^{\,r,r}_{k,l}(-\hat{c},-\hat{d};\hat{e}, \hat{f}) \\
                                         & -
                       \sum_{t=0}^{R+1}\frac{2}{\delta_r^{(t-1)}}\, \delta_{(k,l)(r+1,m)}
        \sum_{\hat{c},\hat{d}} \overline{\psi}^{\, r,r}_{u,v}(\hat{a},\hat{b};\hat{c}, \hat{d})\,  
        \overline{\psi}^{\,r,r}_t(-\hat{c},-\hat{d}; \hat{e},\hat{f}).
\end{split}                                         
\end{equation}
where 
$\overline{\varphi}^{\, r,r}_{k,l}(\hat{a},\hat{b};\hat{e},\hat{f}) = 
[\varphi^{r,r}_{k,l}(\hat{a},\hat{b};\hat{e},\hat{f}) + 
  \varphi^{r,r}_{k.l}(\hat{a},\hat{b};\hat{f},\hat{e})]/2$
and  $m = \max(r+1,t)$.
The second line vanishes, see Eq. (\ref{eq:psikl-psik}), and 
\begin{equation}
\label{eq:Qv2}
\begin{split}
  \sum_{\hat{c},\hat{d}}  \overline{\psi}^{\, r,r}_{u,v}(\hat{a},\hat{b};\hat{c}, \hat{d})\,  
                                        & \overline{\psi}^{\,r,r}_{k,l}(-\hat{c},-\hat{d};\hat{e}, \hat{f})
                                         =   \sum_{\hat{c},\hat{d}}  
                                         \overline{\varphi}^{\,r,r}_{u,v}(\hat{a},\hat{b};\hat{c}, \hat{d})\,  
                                         \overline{\varphi}^{\,r,r}_{k,l}(-\hat{c},-\hat{d};\hat{e}, \hat{f}) \\
                                         & -
                       \sum_{t=0}^{R+1}\frac{2}{\delta_r^{(t-1)}}\, \delta_{(u,v)(r+1,m)}
                       \sum_{\hat{c},\hat{d}} 
                       \overline{\psi}^{\,r,r}_t(\hat{a},\hat{b}; \hat{c}, \hat{d})\,
                       \overline{\varphi}^{\,r,r}_{k,l}(-\hat{c},-\hat{d};\hat{e}, \hat{f})
\end{split}                                         
\end{equation}
By using the the symmetry 
$\overline{\varphi}^{\,r,r}_{k,l}(\hat{c},\hat{d};\hat{a}, \hat{b}) = 
   \overline{\varphi}^{\,r,r}_{k,l}(\hat{a},\hat{b};\hat{c}, \hat{d})$,
and Eq. (\ref{eq:Qv1}),
the second line gives:
\begin{equation}
\label{eq:Qv3}
                     \delta_{(u,v)(k,l)}
                        \sum_{t=0}^{R+1}\frac{2}{\delta_r^{(t-1)}}\, \delta_{(k,l)(r+1,m)}\,
     \overline{\psi}^{\,r,r}_t(\hat{a},\hat{b};\hat{e},\hat{f}),
\end{equation}
where we have used the identity 
$ \delta_{(u,v)(r+1,m)}\, \delta_{(k,l)(r+1,m)} = \delta_{(u,v)(k,l)}\, \delta_{(k,l)(r+1,m)}$.

The sum over $\hat{d}$ in the first line of Eq. (\ref{eq:Qv2})
can be easily performed by exploiting the $\delta_{R+1}$ functions, 
see Eq. (\ref{eq:varphikl}).
Of the four terms consider the one proportional to 
$ \Deltastar_{u; r+1}(\hat{c},0)\, \Deltastar_{v;r+1}(\hat{d},0)
\Deltastar_{k; r+1}(\hat{e},0)\, \Deltastar_{l;r+1}(\hat{f},0)$:
\begin{equation}
\begin{split}
  \sum_{\hat{c},\hat{d}}  \overline{\varphi}^{\, r,r}_{u,v}(\hat{a},\hat{b};\hat{c}, \hat{d})\,  
                                         &\overline{\varphi}^{\, r,r}_{k,l}(-\hat{c},-\hat{d};\hat{e}, \hat{f}) \Rightarrow \\
                    &\frac{1}{4p_0^2}\Bigl[
                    \sum_{\hat{c}} \pstar_r(\hat{a}+\hat{c},0) 
                         \bigl[ \pstar_r(\hat{c},\hat{e}) + \pstar_r(\hat{c},\hat{f})\bigr] \\
                         &\phantom{=}\times
                         \Deltastar_{u,r+1}(\hat{c},0) \Deltastar_{v,r+1}(\hat{e}+\hat{f}-\hat{c},0)
                         \Deltastar_{k,r+1}(\hat{e},0) \Deltastar_{l,r+1}(\hat{f},0) \\
&  \phantom{===}       + (\hat{a}\leftrightarrow\hat{b})
                    \Bigr] \,                                                        
                    \delta_{R+1}(\hat{a}+\hat{b}+\hat{e}+\hat{f},0).
\end{split}  
\end{equation}
For $u,v\geq r+1$ the $\hat{c}$ into 
$\Deltastar_{u,r+1}(\hat{c},0) \Deltastar_{v,r+1}(\hat{e}+\hat{f}-\hat{c},0)$ can be replaced 
by $\hat{a}$,
and the sum over $\hat{c}$ evaluated with the help of Eq. (\ref{eq:pstar_pstar}):
\begin{equation}
\begin{split}
  \sum_{\hat{c},\hat{d}}  &\overline{\varphi}^{\, r,r}_{u,v}(\hat{a},\hat{b};\hat{c}, \hat{d})\,  
                                         \overline{\varphi}^{\, r,r}_{k,l}(-\hat{c},-\hat{d};\hat{e}, \hat{f}) \Rightarrow \\
                    &\phantom{==} \frac{1}{4p_0}\Bigl[
                         \bigl[ \pstar_r(\hat{a}+\hat{e},0) + \pstar_r(\hat{a}+\hat{f},0)\bigr]\\
                         &\phantom{==}\times
                         \Deltastar_{u,r+1}(\hat{a},0) \Deltastar_{v,r+1}(\hat{e}+\hat{f}-\hat{a},0)
                         \Deltastar_{k,r+1}(\hat{e},0) \Deltastar_{l,r+1}(\hat{f},0) \\
&  \phantom{====}       + (\hat{a}\leftrightarrow\hat{b})
                    \Bigr] \,                                                        
                    \delta_{R+1}(\hat{a}+\hat{b}+\hat{e}+\hat{f},0).
\end{split}  
\end{equation}
The (unnormalized) projector $\pstar_{r}(\hat{a}+\hat{e},0)$ implies that $\hat{a}+\hat{e}=0$ on 
levels $r+1,\dotsc,R+1$, 
and hence $u=k$ and $v=l$, otherwise the $\Deltastar$ terms vanish. 
Similarly $\pstar_{r}(\hat{a}+\hat{f},0)$ implies  $u=l$ and $v=k$.
Then 
\begin{equation}
\begin{split}
  \sum_{\hat{c},\hat{d}} & \overline{\varphi}^{\, r,r}_{u,v}(\hat{a},\hat{b};\hat{c}, \hat{d})\,  
                                         \overline{\varphi}^{r,r}_{k,l}(-\hat{c},-\hat{d};\hat{e}, \hat{f}) \Rightarrow \\
                    & \phantom{=} \frac{1}{4p_0}\Bigl[
                         \bigl[\delta_{u,k}\,\delta_{v,l}\,\pstar_r(\hat{a}+\hat{e},0) + 
                                  \delta_{u,l}\,\delta_{v,k}\, \pstar_r(\hat{a}+\hat{f},0)\bigr]
%                                  &\phantom{====}\times
                         \Deltastar_{k,r+1}(\hat{e},0) \Deltastar_{l,r+1}(\hat{f},0) \\
&  \phantom{==========}       + (\hat{a}\leftrightarrow\hat{b})
                    \Bigr] \,                                                        
                    \delta_{R+1}(\hat{a}+\hat{b}+\hat{e}+\hat{f},0) \\
                   & = \frac{1}{4p_0}\,
                    \delta_{(u,v)(k,l)}\,
                    \Bigl[
                         \pstar_r(\hat{a}+\hat{e},0) +  \pstar_r(\hat{a}+\hat{f},0)\Bigr]\,
                         \Deltastar_{k,r+1}(\hat{e},0) \Deltastar_{l,r+1}(\hat{f},0) \\
                         &\phantom {===========} \times
                    \delta_{R+1}(\hat{a}+\hat{b}+\hat{e}+\hat{f},0)          
\end{split}  
\end{equation} 
\if 0
and rearranging the terms: 
\begin{equation}
\label{eq:Qv4}
\begin{split}
  \sum_{\hat{c},\hat{d}}  \varphi^{r,r}_{u,v}(\hat{a},\hat{b};\hat{c}, \hat{d})\,  
                                         &\varphi^{r,r}_{k,l}(-\hat{c},-\hat{d};\hat{e}, \hat{f}) \Rightarrow \\
                    &\frac{1}{4p_0}\,
                    \delta_{(u,v)(k,l)}\,
                    \Bigl[
                         \pstar_r(\hat{a}+\hat{e},0) +  \pstar_r(\hat{a}+\hat{f},0)\Bigr]\\
                         &\phantom {==} \times
                         \Deltastar_{k,r+1}(\hat{e},0) \Deltastar_{l,r+1}(\hat{f},0) \,
                    \delta_{R+1}(\hat{a}+\hat{b}+\hat{e}+\hat{f},0)
\end{split}  
\end{equation}
\fi
The contribution from the other three terms is obtained by suitable exchanges of 
$u$ and $v$ and $k$ and $l$. Summing up all contributions leads to: 
\begin{equation}
\label{eq:Qv4-s}
  \sum_{\hat{c},\hat{d}}  \overline{\varphi}^{\, r,r}_{u,v}(\hat{a},\hat{b};\hat{c}, \hat{d})\,  
                                         \overline{\varphi}^{\, r,r}_{k,l}(-\hat{c},-\hat{d};\hat{e}, \hat{f}) =
                    \delta_{(u,v)(k,l)}\, \overline{\varphi}^{\,r,r}_{k,l}(\hat{a},\hat{b};\hat{e}, \hat{f}).
\end{equation}

Then, from Eqs. (\ref{eq:Qv2}), (\ref{eq:Qv3}) and (\ref{eq:Qv4-s}):
\begin{equation}
%\label{eq:Qv2}
  \sum_{\hat{c},\hat{d}}  \overline{\psi}^{\,r,r}_{u,v}(\hat{a},\hat{b};\hat{c}, \hat{d})\,  
                                        \overline{\psi}^{\,r,r}_{k,l}(-\hat{c},-\hat{d};\hat{e}, \hat{f})
     = \delta_{(u,v)(k,l)}\, \overline{\psi}^{\, r,r}_{k,l}(\hat{a},\hat{b};\hat{e}, \hat{f}).
\end{equation}
By repeating the calculation for $r\not=s$ it is easy to see that the sum vanishes because of the 
$\delta_{r,s}$ terms  in Eq. (\ref{eq:Qv1}) and Eq. (\ref{eq:pstar_pstar}). 
The use of the symmetry 
$\overline{\psi}^{\, r,r}_{k,l}(\hat{a},\hat{b};\hat{c}, \hat{d}) = 
   \overline{\psi}^{\, r,r}_{k,l}(\hat{c},\hat{d};\hat{a}, \hat{b})$,
and rearranging the indexes, completes the proof of Eq. (\ref{eq:s-psikl-s-psikl}).   (Q.E.D)

\section{Eigenvalues of 1RSB fluctuations in the Spherical Model.}
\label{app:Fluct1RSB}

From the general results of Section \ref{sec:SphMod} valid for any $R$, in the case of
one replica symmetry breaking step $R=1$ the eigenvalues in the Replicon sector are
($r = 0, 1$), see Eq. (\ref{eq:RRFTsp}):

\begin{equation}
 A^{1,1}_{\hat{2},\hat{2}} = -\Lambda'(Q_1) + \frac{1}{(Q_{\hat{2}})^2}
\end{equation}
\begin{equation}
 A^{0,0}_{\hat{2},\hat{2}} = -\Lambda'(Q_0) + \frac{1}{(Q_{\hat{2}})^2}
\end{equation}
\begin{equation}
 A^{0,0}_{\hat{1},\hat{1}} = -\Lambda'(Q_0) + \frac{1}{(Q_{\hat{1}})^2}
\end{equation}
\begin{equation}
 A^{0,0}_{\hat{1},\hat{2}} = A^{0,0}_{\hat{2},\hat{1}} =
 -\Lambda'(Q_0) + \frac{1}{Q_{\hat{1}}Q_{\hat{2}}}
\end{equation}
where
\begin{equation}
 Q_{\hat{2}} = 1 - Q_1, \qquad
 Q_{\hat{1}} = 1 - Q_1 + p_1 (Q_1 - Q_0)
\end{equation}
since $Q_{2}=1$ from the spherical constraint.
These correspond to the eigenvalues $\Lambda^{(1)}_1$, $\Lambda^{(1)}_0$, 
$\Lambda^{(3)}_0$ and $\Lambda^{(2)}_0$ of Ref. \cite{CriSom92}, respectively.

The eigenvalue in the LA sector are given by the solution of the eigenvalue 
equation (\ref{eq:eigenLA}), which for $R=1$ reads
\begin{equation}
\label{eq:LA1RSB}
\left\{
\begin{split}
 \Bigl[ A^{0,0}_{\hat{1},\hat{m}} + \frac{1}{4}\delta^{(k-1)}_0 A^{0,0}_{\hat{k}}  \Bigr]\, \phi^0_{\hat{k}}
    + \frac{1}{4}\delta^{(k-1)}_1 A^{0,1}_{\hat{k}}\, \phi^1_{\hat{k}} &= \Lambda\, \phi^0_{\hat{k}}\\
      \frac{1}{4}\delta^{(k-1)}_1 A^{1,0}_{\hat{k}}\, \phi^0_{\hat{k}}
+ \Bigl[ A^{1,1}_{\hat{2},\hat{2}} + \frac{1}{4}\delta^{(k-1)}_1 A^{1,1}_{\hat{k}}  \Bigr]\, \phi^1_{\hat{k}}
  &= \Lambda\, \phi^1_{\hat{k}},
\end{split}
\right.
\end{equation}
with $k=0,1,2$ and $m = \max(1,k)$.

From Eq. (\ref{eq:LARFTsp}) for $k=0$ we have:
\begin{equation}
 A^{0,0}_{\hat{0}} =  
A^{0,1}_{\hat{0}} =  
 2\, p_0\, (Q^{-1}_0)^2 + 4 \frac{Q^{-1}_0}{Q_{\hat{1}}},
\end{equation}
\begin{equation}
 A^{1,1}_{\hat{0}} =  
 2\, p_0\, (Q^{-1}_0)^2 + 2 p_1 \bigl[ (Q^{-1}_1)^2 - (Q^{-1}_0)^2 \bigr]
+  4 \frac{Q^{-1}_1}{Q_{\hat{2}}}.
\end{equation}
By transforming these expressions into the $(A,B,C,m)$ notation of Ref. \cite{CriSom92} by using:
\begin{equation}
 A = \frac{1}{Q_{\hat{2}}}, \quad
 B = - \frac{Q_1 - Q_0}{Q_{\hat{2}} Q_{\hat{1}}}, \quad
 C = - \frac{Q_0}{Q_{\hat{1}} Q_{\hat{0}}}, \quad
 m = p_1,
\end{equation}
and $n = p_0$, and replacing them into the eigenvalue equation (\ref{eq:LA1RSB}),
a straightforward calculation shows that for $k=0$ one recovers the eigenvalue
$\Lambda^{(3)}_{3,4}$ of Ref. \cite{CriSom92}.

For $k=1$ we have:
 \begin{equation}
 A^{0,0}_{\hat{1}} =  
A^{0,1}_{\hat{1}} =  
    4 \frac{Q^{-1}_0}{Q_{\hat{1}}},
\end{equation}
\begin{equation}
 A^{1,1}_{\hat{1}} =  
   2 p_1 \bigl[ (Q^{-1}_1)^2 - (Q^{-1}_0)^2 \bigr]
+  4 \frac{Q^{-1}_1}{Q_{\hat{2}}}.
\end{equation}
and in this case one recovers, with a bit of work, the eigenvalues 
$\Lambda^{(3)}_{1,2}$.

Finally for $k=2$ 
 \begin{equation}
 A^{0,0}_{\hat{2}} =  
A^{0,1}_{\hat{1}} =  
    4 \frac{Q^{-1}_0}{Q_{\hat{2}}},
\end{equation}
\begin{equation}
 A^{1,1}_{\hat{2}} = 4 \frac{Q^{-1}_1}{Q_{\hat{2}}}.
\end{equation}

For a mistake unseen at the proof reading stage  for this case in Ref. \cite{CriSom92}
only the relevant eigenvalue for $Q_0=0$ is reported.
Indeed, in the $(A,B,C,m)$ notation the eigenvalue
equation reads:
\begin{equation}
\left\{
\begin{split}
 \Bigl[ -\Lambda'(Q_0) + A(A+mB) + (n-m) AC  \Bigr]\, \phi^0_{\hat{k}}
    + (m-2) AC \, \phi^1_{\hat{k}} &= \Lambda\, \phi^0_{\hat{k}}\\
      (n-m) AC \, \phi^0_{\hat{k}}
+ \Bigl[ -\Lambda'(Q_1) + A^2 + (m-2) A (B+C)  \Bigr]\, \phi^1_{\hat{k}}
  &= \Lambda\, \phi^1_{\hat{k}},
\end{split}
\right.
\end{equation}
so that the eigenvalues have a form similar to that of $\Lambda^{(3)}_{1,2}$ or 
$\Lambda^{(3)}_{3,4}$, solution of a second order equation.
However if $Q_0=0$ then $C=0$ and the eigenvalue equation becomes diagonal 
with eigenvalues
\begin{equation}
 \Lambda = -\Lambda'(0) + A(A+mB),
\end{equation}
and
\begin{equation}
 \Lambda = -\Lambda'(Q_1) + A\bigl[A + (m-2)B \bigr],
\end{equation}
which is the eigenvalue $\Lambda^{(2)}_1$ [for $C=0$] reported in Ref. \cite{CriSom92}.
For the model studied in \cite{CriSom92} $\Lambda'(0) = 0$ and the first eigenvalue is 
irrelevant for the stability since $A(A+mB) = 1/Q_{\hat{2}}Q_{\hat{1}}$.
Note also a misprints in the degeneracy of $\Lambda^{(2)}_0$. The correct degeneracy is indeed
$(n/m)(m-1)(n/m-2)$.

%%%%%%%%%%%%%%%%%%%%%%%%%%%%%%%%%%%%%%%%%%%%%%%%%


\newpage
\begin{thebibliography}{99}

\bibitem{SheKir75}
  D. Sherrington, S. Kirkpatrick,
  Phys. Rev. Lett. {\bf 35} 1792 (1975)

\bibitem{SheKir78}
  S. Kirkpatrick, D. Sherrington,
  Phys. Rev. B {\bf 17}, 4384  (1978)

\bibitem{Parisi79}
  G. Parisi,
  Phys. Rev. Lett. {\bf 43}, 1754  (1979) 

\bibitem{Parisi80}
  G. Parisi, 
  J. Phys. A {\bf 13}, 1101  (1980)

%\bibitem{Franzetal13}
%  S. Franz, H. Jacquin, G. Parisi, P. Urbani, F. Zamponi
%  J. Chem. Phys. {\bf 138}, 12A540 (2013) 

\bibitem{BraMoo86}
  A. Bray and M. A. Moore, 
  in {\it Heidelberg Colloquium on Glassy dynamics and optimizations},
  L. Van Hemmen and I. Morgensten, eds., 
  Springer-Verlag, 1986.

\bibitem{MooBra11}
  M. A. Moore, A. J. Bray,
  Phys. Rev. B {\bf 83},  224408 (2011) 

\bibitem{ParTem12}
  G. Parisi, T. Temesv\'ari
  Nucl. Phys. B {\bf 858}, 293  (2012) 

\bibitem{DeDomCar96}
     C. De Dominicis, D.M. Carlucci,
     C. R. Acad. Sci Paris, t. 323, S\'erie II b, 263 (1996)

\bibitem{DeDomCarTem97}
     C. De Dominicis, D. M. Carlucci, T. Temesv\'ari,
     J. Phys. I France {\bf 7}, 105 (1997)

\bibitem{CarDeDom97}
      D. M. Carlucci, C. De Dominicis,
      arXiv:cond-mat/9709200 (1997)

\bibitem{DeDomKonTem98}
    C. De Dominicis, I. Kondor,  T. Temesv\'ari,
   {\sl Beyond the Sherrington-Kirkpatrick Model}
   (World Scientific, 1998), vol. 12 
    of Series on Directions in Condensed Matter Physics, 
    p. 119, cond-mat/9705215.

\bibitem{CriSom92}
     A. Crisanti, H.-J. Sommers,
     Z. Phys. B {\bf 87}, 341 (1992)

\bibitem{Temal94}
     T. Temesv\'ari, C. De Dominicis, I. Kondor,
     J. Phys. A {\bf 27}, 7569 (1994)

\bibitem{CriLeu04}
    A. Crisanti, L. Leuzzi
    Phys. Rev. Lett. {\bf 93}, 217203 (2004)

\bibitem{CriLeu06}
   A. Crisanti, L. Leuzzi
   Phys. Rev. B {\bf 73}, 014412 (2006)

\bibitem{CriLeu07}
    A. Crisanti, L. Leuzzi
   Phys. Rev. B {\bf 76}, 184417 (2007)

\bibitem{Sunetal12}
     Y. Sun, A. Crisanti, F. Krzakala, L. Leuzzi, L. Zdeborov‡
    J.  Stat/ Mech. 2012, P07002 (2012)

\bibitem{CriLeu13}
    A. Crisanti, L. Leuzzi, 
    Nucl. Phys. B {\bf 870} [FS], 176 (2013)
    
\end{thebibliography}
\end{document}